\documentclass[fleqn,usenatbib]{mnras}

\usepackage{newtxtext}
\usepackage{xcolor}

\usepackage[T1]{fontenc}

\DeclareRobustCommand{\VAN}[3]{#2}
\let\VANthebibliography\thebibliography
\def\thebibliography{\DeclareRobustCommand{\VAN}[3]{##3}\VANthebibliography}

\usepackage{graphicx}	
\usepackage{amsmath}	
\usepackage{amssymb}	

\title[Metal-Poor AGNs]{On the AGN Nature of Broad Balmer Emission in Four Low-Redshift Metal-Poor Galaxies}

\author[C. J. Burke et al.]{
Colin J. Burke,$^{1,2}$\thanks{E-mail: colinjb2@illinois.edu}
Xin Liu,$^{1,2}$
Yu-Ching Chen,$^{1,2}$
Yue Shen,$^{1,2}$
and Hengxiao Guo$^{1,2,3}$
\\
$^{1}$Department of Astronomy, University of Illinois at Urbana-Champaign, 1002 West Green Street, Urbana, IL 61801, USA\\
$^{2}$National Center for Supercomputing Applications, 1205 West Clark Street, Urbana, IL 61801, USA\\
$^{3}$Department of Physics and Astronomy, 4129 Frederick Reines Hall, University of California, Irvine, CA, 92697-4575, USA\\
}

\date{Accepted XXX. Received YYY; in original form ZZZ}

\pubyear{2021}

\begin{document}
\label{firstpage}
\pagerange{\pageref{firstpage}--\pageref{lastpage}}
\maketitle

\begin{abstract}
We report on continued, $\sim$15 year-long, broad Balmer emission lines in three metal-poor dwarf emission-line galaxies selected from Sloan Digital Sky Survey spectroscopy. The persistent luminosity of the broad Balmer emission indicates the galaxies are active galactic nuclei (AGNs) with virial black hole masses of $\sim 10^{6.7}{-10^{7.0}\ M_{\odot}}$. The lack of observed hard X-ray emission and the possibility that the Balmer emission could be due to a long-lived stellar transient motivated additional follow-up spectroscopy. We also identify a previously-unreported blueshifted narrow absorption line in the broad H$\alpha$ feature in one of the AGNs, indicating an AGN-driven outflow with hydrogen column densities of order $10^{17}$ cm$^{-2}$. We also extract light curves from the Catalina Real-Time Transient Survey and the Zwicky Transient Facility. We detect probable AGN-like variability in three galaxies, further supporting the AGN scenario. This also suggests the AGNs are not strongly obscured. This sample of galaxies are among the most metal-poor which host an AGN ($Z=0.05$ -- $0.16\ Z_\odot$). We speculate they may be analogues to seed black holes which formed in unevolved galaxies at high redshift. Given the rarity of metal-poor AGNs and small sample size available, we investigate prospects for their identification in future spectroscopic and photometric surveys.

\end{abstract}




\section{Introduction}

In a search for emission-line galaxies (ELGs) in Sloan Digital Sky Survey (SDSS) DR5 spectra, \citet{Izotov2007} identified four metal-poor ELGs with extremely luminous broad Balmer emission. \citet{Izotov2008} describe the four galaxies in detail (hereafter the IT08 sample). The criteria of their search involved selection of spectra with $[\text{O}~\textsc{iii}]$ $\lambda4363$ line detections to directly measure the element abundances, and exclusion of obvious high-metallicity AGNs. They were left with a sub-sample of about 10,000 ELGs, and noticed four metal-poor ELGs with broad H$\alpha$ emission. Their broad H$\alpha$ luminosities range from $3\times10^{41}$ to $2\times10^{42}$ erg s$^{-1}$ with FWHM of about 1500 -- 2000 km s$^{-1}$. Further observations demonstrated the broad emission in all four galaxies is long-lived (10 -- 13 years; \citealt{Izotov2008,Simmonds2016}).

There are several mechanisms which can produce broad Balmer lines in ELGs. Stellar winds due to massive, eruptive stars (Wolf-Rayet, luminous blue variables) can produce broad H$\alpha$ luminosities up to $10^{40}$ erg s$^{-1}$. But according to \cite{Izotov2008}, anything larger must be attributed to shocks from supernovae (SNe) or the broad line region of an active galactic nucleus (AGN). In the former scenario, the broad emission should fade away after several years \citep{Baldassare2016}. \citet{Izotov2008} favor the AGN interpretation for the IT08 galaxies. However, no hard X-ray emission was detected in \emph{Chandra} observations \citep{Simmonds2016}, in contrast to the naive expectation from the AGN interpretation\footnote{Three soft photons were detected in SDSS J1047, consistent with a marginal detection of an underlying X-ray binary population.}.

One additional galaxy identified by \citet{Izotov2007} was the blue compact dwarf PHL 293B \citep{Izotov2009,Izotov2011}. Its broad H$\alpha$ luminosity was $\sim 1\times10^{39}$ erg s$^{-1}$ in 2001 and relatively constant until it was observed to fade after 2011 \citep{Burke2020b,Allan2020}. A similar persistent transient, SDSS1133, was identified by \citet{Koss2014}. Motivated by the possibility that the broad Balmer emission lines of the IT08 sample could be attributed to extraordinarily long-lived SNe or stellar eruption powered by interaction with a dense circumstellar medium, or an unknown stellar process, we obtained follow-up extended-baseline spectroscopy with the Gemini Observatory in 2020 for three of the four IT08 ELGs.

We find the broad emission is still present in our three targets for over 15 years since the first SDSS observations, along with long-duration photometric variability in three of the galaxies. This work extends the previous spectral baselines for the three observed galaxies by 4 to 5 years. This indicates the AGN interpretation remains the most plausible scenario. Consistent with the lack of strong X-ray emission, we find no radio detections in archival Very Large Array (VLA) imaging. The IT08 ELGs lie at least 1--2 dex below the expected X-ray emission of more typical AGNs. In one source, we identify a previously-unreported narrow absorption feature in the broad H$\alpha$ emission component, indicating hydrogen column densities along the line-of-sight of $N_{\text{HI}} \approx 2 \times10^{17}$ cm$^{-2}$. We interpret this as evidence for an AGN-driven outflow.

We conclude the IT08 ELGs are in fact a rare new class of X-ray/radio-weak dwarf AGNs with black hole masses of $10^{6.7}{-10^{7.0}\ M_{\odot}}$. The host galaxy stellar masses of the IT08 AGNs estimated from the broadband spectral energy distribution (SED) are in the range $\sim 10^{8.4} - 10^{10.0} M_{\odot}$, typical of metal-poor compact emission line galaxies \citep{Cardamone2009}. There has been considerable interest in identifying such systems in the local universe, with the motivation that they may be analogues to primordial supermassive black hole (SMBH) seeds that formed at high redshift \citep{Volonteri2008,Greene2012}. \citet{Mezcua2019} reminds us that typical low-$z$ dwarf galaxies have undergone growth via multiple mergers throughout their history. Therefore, metal-poor dwarf AGNs should be more pristine and unevolved analogues to SMBH seeds. However, dwarf AGNs are difficult to identify using traditional techniques because of their low AGN luminosities and because low-metallicity galaxies shift toward the star-forming region of the BPT diagram \citep{Groves2006}. Motivated by this, we investigate prospects for identifying IT08-like metal-poor AGNs in future spectroscopic surveys.

This work is organised as follows. In \S\ref{sec:observations}, we describe our new Gemini spectroscopic observations, light curve data, and archival VLA imaging. In \S\ref{sec:results}, we describe the results of our analysis. In \S\ref{sec:prospects}, we discuss prospects for identifying similar AGNs in future spectroscopic surveys. In \S\ref{sec:conclusions}, we summarize our results and conclude. A concordance $\Lambda$CDM cosmology with $\Omega_m = 0.3$, $\Omega_{\Lambda} = 0.7$, and $H_{0}=70$ km s$^{-1}$ Mpc$^{-1}$ is assumed throughout.

\section{Observations and Data Reduction}
\label{sec:observations}

\begin{figure*}
	\includegraphics[width=\textwidth]{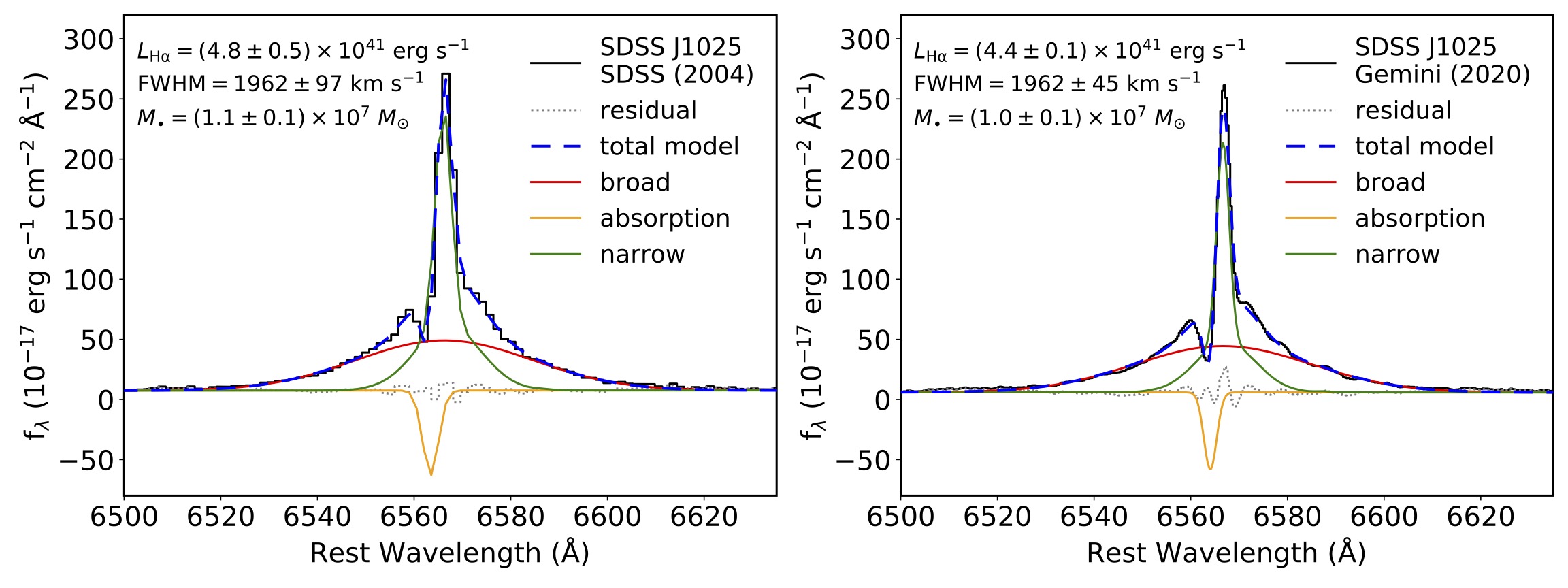}
	\includegraphics[width=\textwidth]{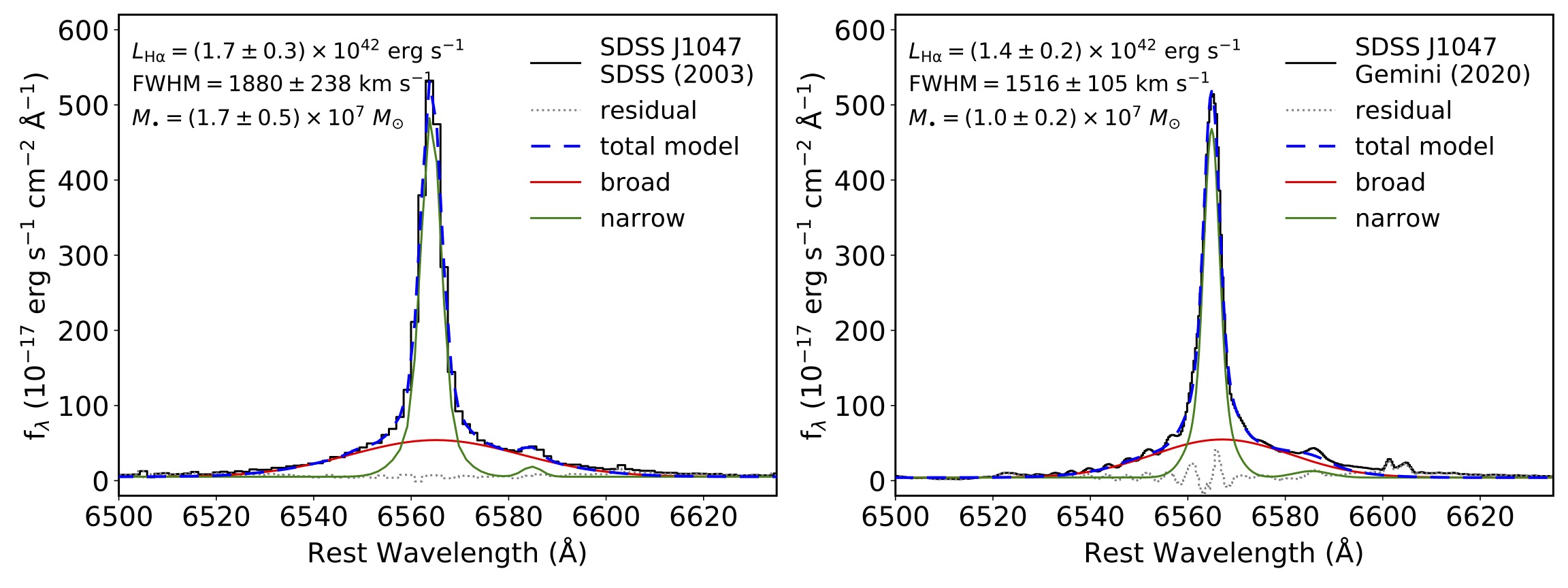}
	\includegraphics[width=\textwidth]{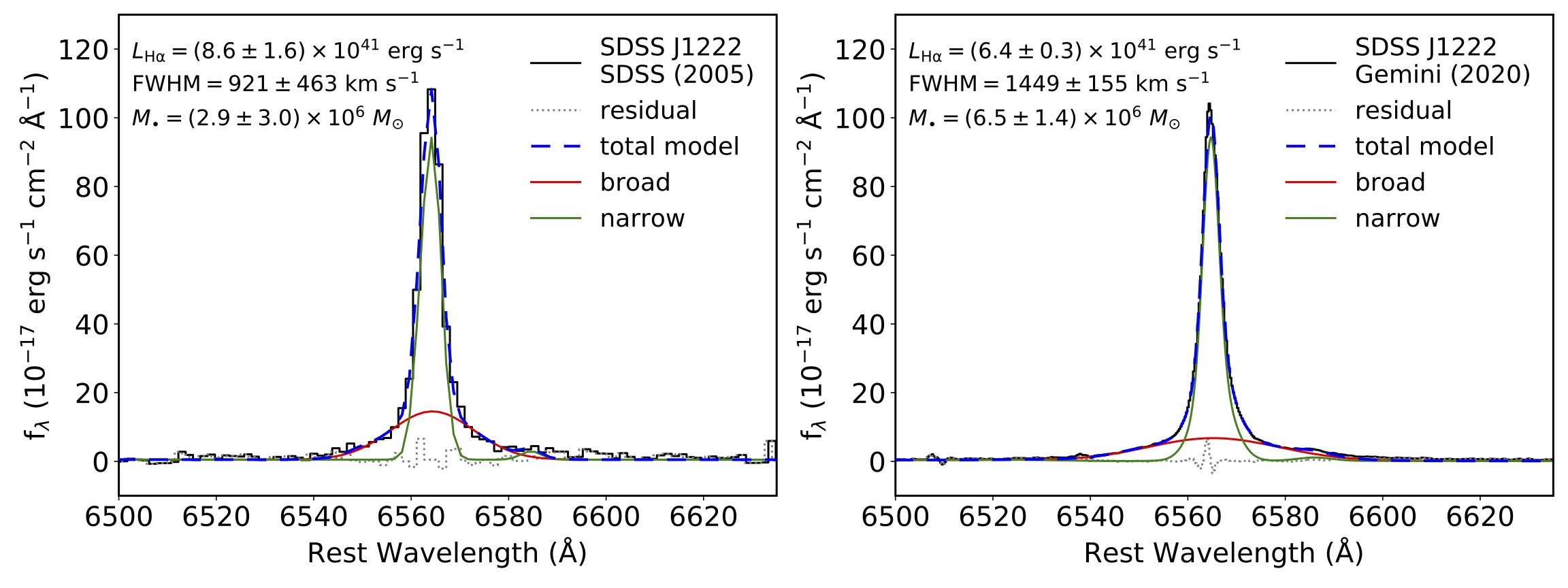}
    \caption{H$\alpha$ spectrum of SDSS J1025 (\emph{top row}), SDSS J1047 (\emph{middle row}), and SDSS J1222 (\emph{bottom row}). from SDSS taken from 2003 -- 2005 (\emph{left}) and Gemini taken in 2020 (\emph{right}). The data are shown in black, the total model in blue, the narrow Gaussian component in green, and the broad Gaussian component in red. The absorption line in SDSS J1025 is shown in orange. The residual is shown as a light-grey dotted line. The luminosity and FWHM refer to the broad component shown in red. Only statistical uncertainties are shown. The narrow absorption feature in the spectrum of SDSS J1025 is clearly visible in high-resolution Gemini spectrum.}
    \label{fig:spectra}
\end{figure*}

\begin{figure*}
	\includegraphics[width=\textwidth]{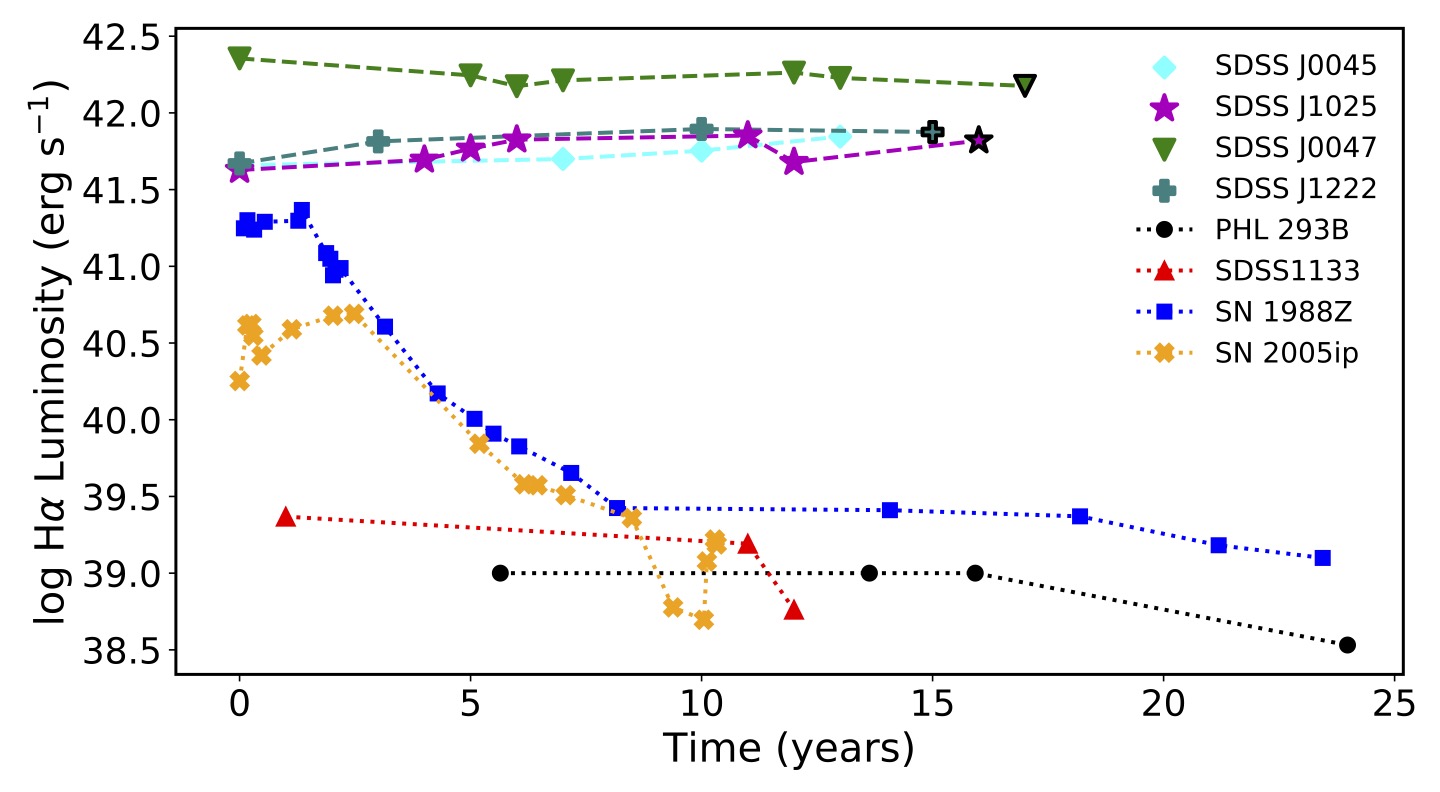}
    \caption{Broad H$\alpha$ luminosity versus time for the IT08 sample (dashed lines with diamonds, stars, inverted triangles, plus symbols) compared to PHL 293B (black circles), SDSS1133 (red triangles), SN 1988Z (blue squares), and SN 2005ip (orange crosses). Our new Gemini data are shown with black borders. Uncertainties are dominated by systematic uncertainties, which are difficult to quantify exactly, but are typically $\pm$20 per cent.}
    \label{fig:Halphaluminosity}
\end{figure*}

\subsection{Gemini Spectra}

We obtained new Gemini long-slit spectroscopy in March -- May 2020 using the GMOS instrument on Gemini-North (GN-2020A-FT-204; PI Burke) for three of the four IT08 galaxies: SDSS J1047+0739, SDSS J1222+3602, and SDSS J1025+1402. We targeted the H$\alpha$ line with the R831 grating and a 1$^{\prime\prime}$ slit width.


We followed the GMOS Cookbook for the reduction of long-slit spectra with \textsc{PyRAF} \footnote{\url{http://ast.noao.edu/sites/default/files/GMOS\_Cookbook}}. The steps include bias subtraction, flat-field correction, wavelength calibration, cosmic ray rejection, and flux calibration using the flux standard star Feige 66. Our flux-calibrated Gemini spectra of SDSS J1047 and SDSS J1222 showed variations in the narrow-line fluxes relative to the SDSS spectra due to seeing and aperture variations compared to the 3$^{\prime\prime}$ SDSS fibre. To mitigate aperture effects and allow for a fair comparison of the broad Balmer luminosities, we normalised the Gemini spectrum to the SDSS flux-levels using the peak narrow H$\alpha$ flux levels. 

We fit the H$\alpha$ region of each spectrum using multi-component Gaussians using the \textsc{PyQSOFit} code \citep{Guo2018,Shen2019}. We fit a continuum and Gaussian emission/absorption lines within user-defined windows and constraints on their widths. The continuum is modeled as a blue power-law plus a 3rd-order polynomial for reddening. The total model is a linear combination of the continuum and single or multiple Gaussians for the emission lines. Since uncertainties in the continuum model may induce subtle effects on measurements for weak emission lines, we first perform a global fit to the emission-line free region to better quantify the continuum.

We then fit multiple Gaussian models to the continuum-subtracted spectra around the H$\alpha$ emission line region locally. We use two narrow Gaussians and one broad Gaussian to model the Gemini H$\alpha$ emission lines of our spectra. For the SDSS J1025, we add a narrow Gaussian with a negative amplitude to model the absorption line. We define narrow Gaussians as having $\rm{FWHM}<800$~km~s$^{-1}$. The narrow and broad line centroids are fit within a window of $\pm65$~\AA\ and $\pm100$~\AA, respectively. We use 50 Monte Carlo simulations to estimate the uncertainty in the line measurements.

\subsection{Light Curves}

\begin{figure*}
    \includegraphics[width=0.49\textwidth]{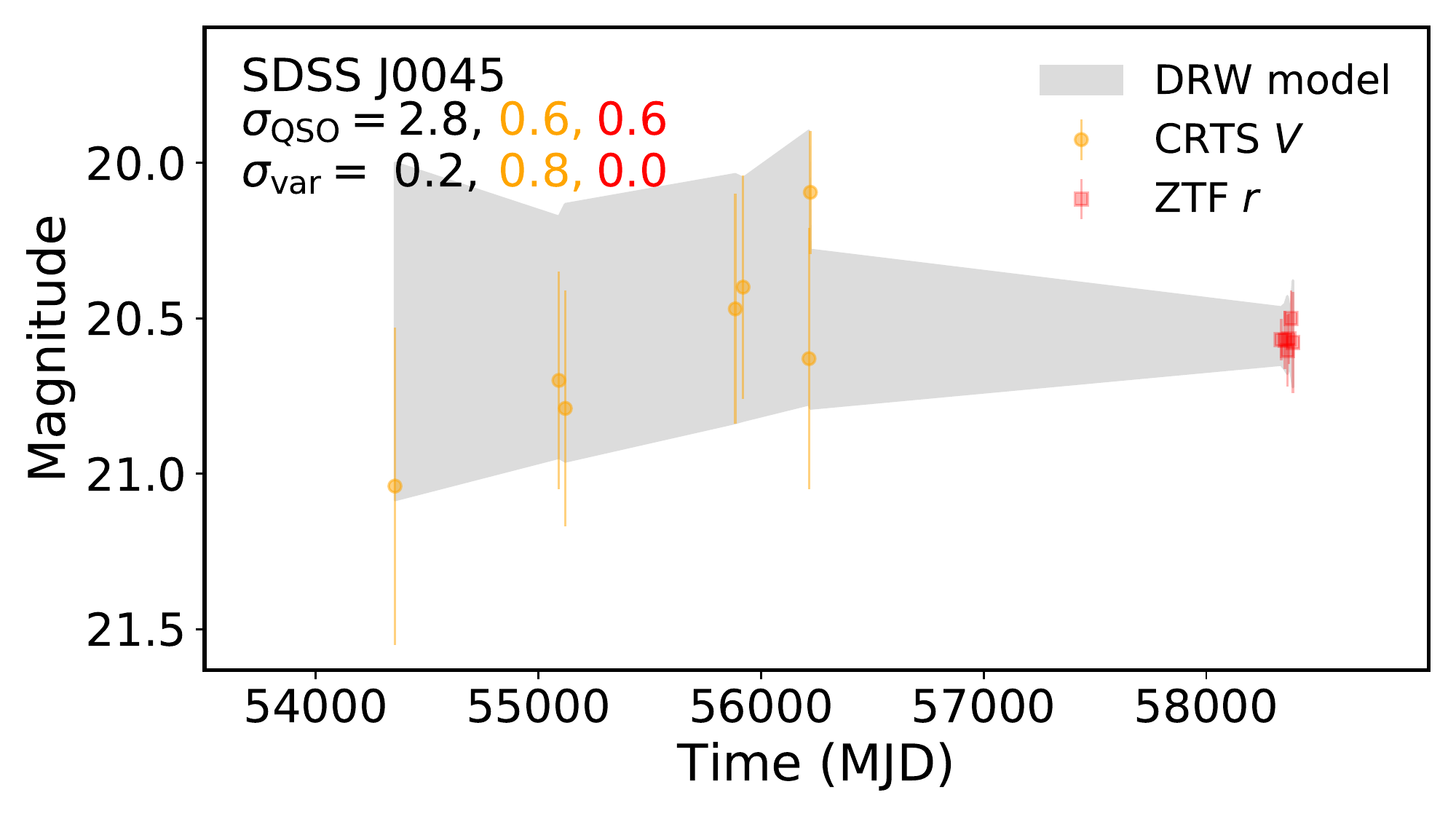}
	\includegraphics[width=0.49\textwidth]{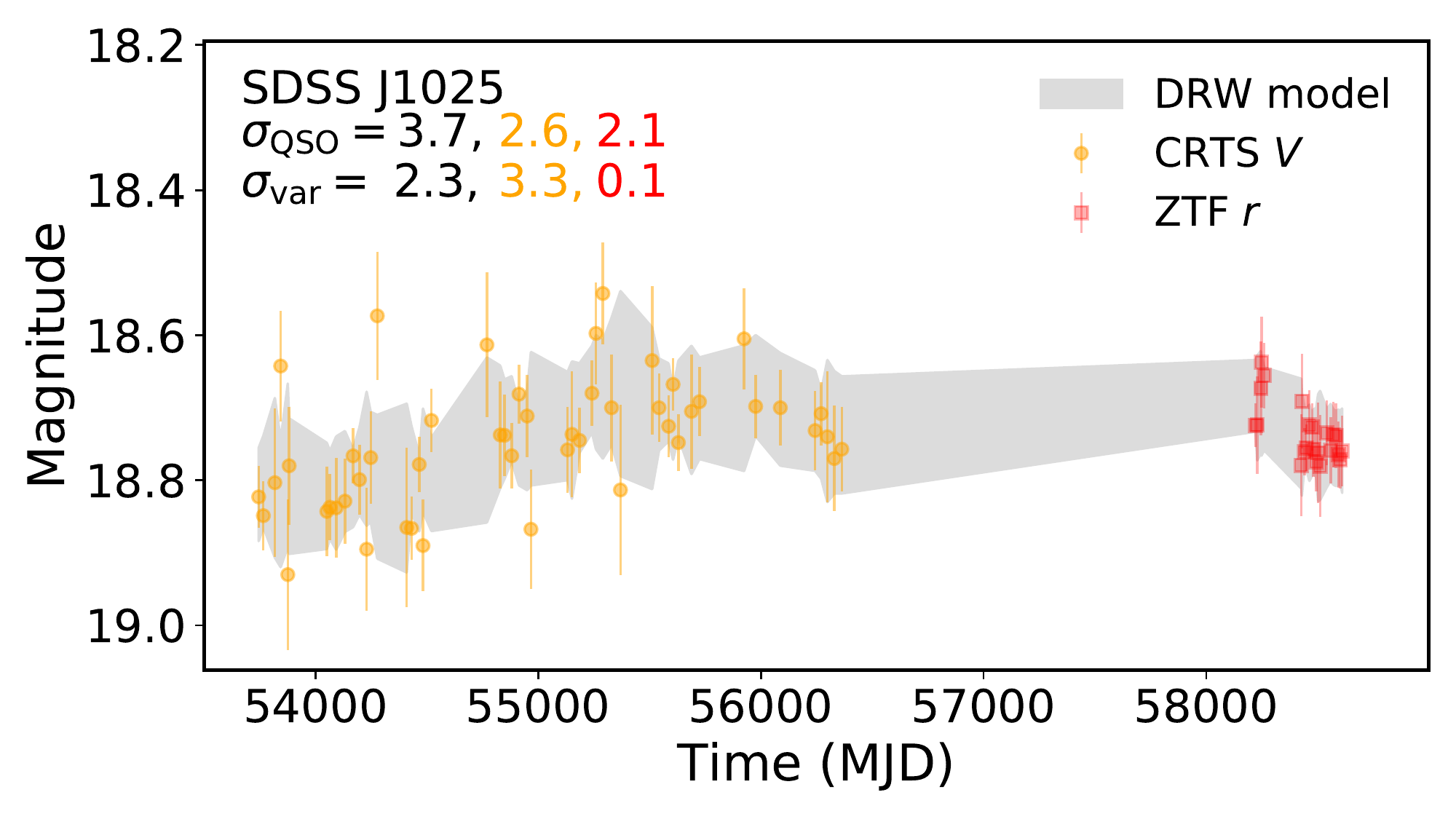}
	\includegraphics[width=0.49\textwidth]{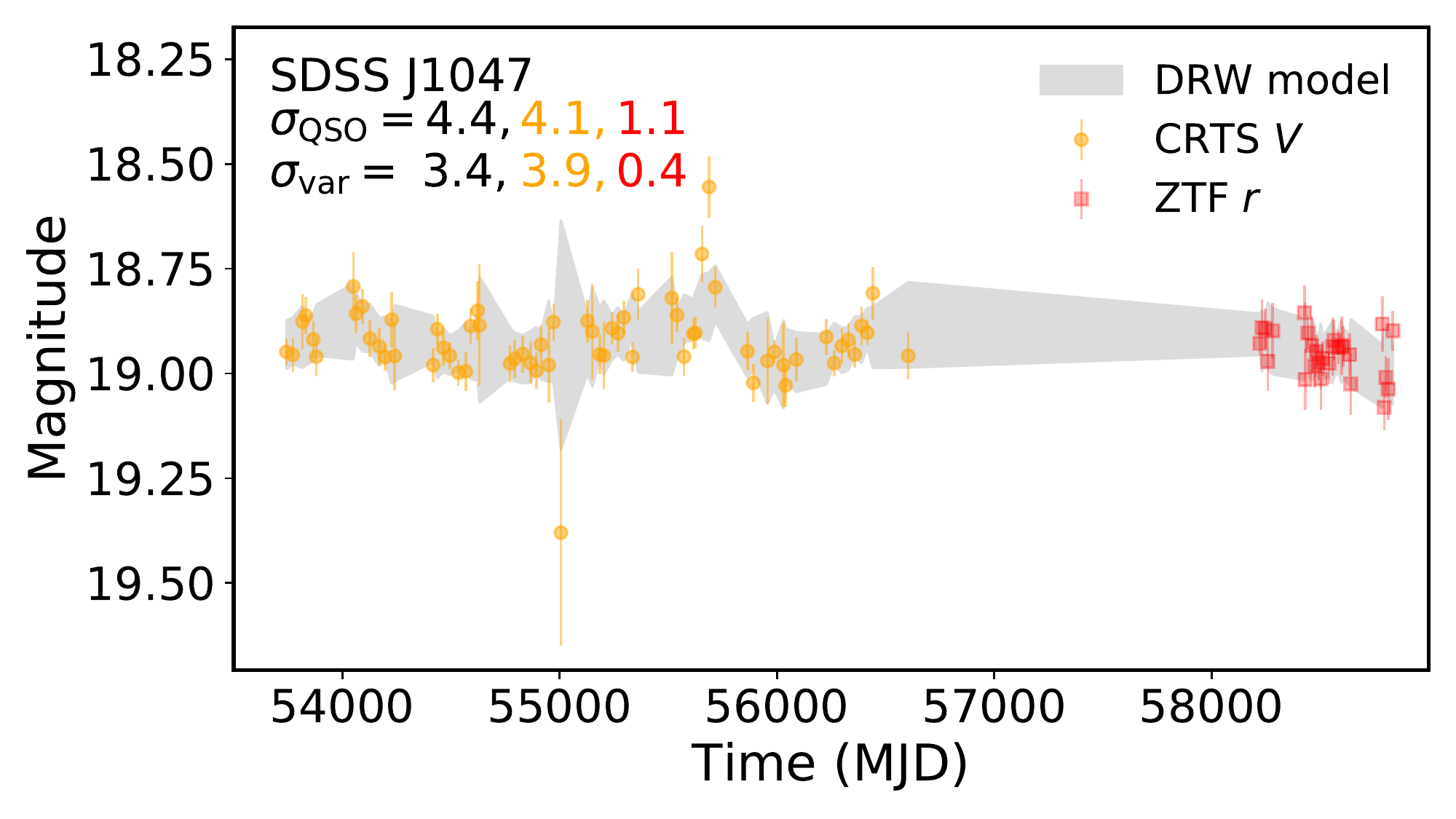}
	\includegraphics[width=0.49\textwidth]{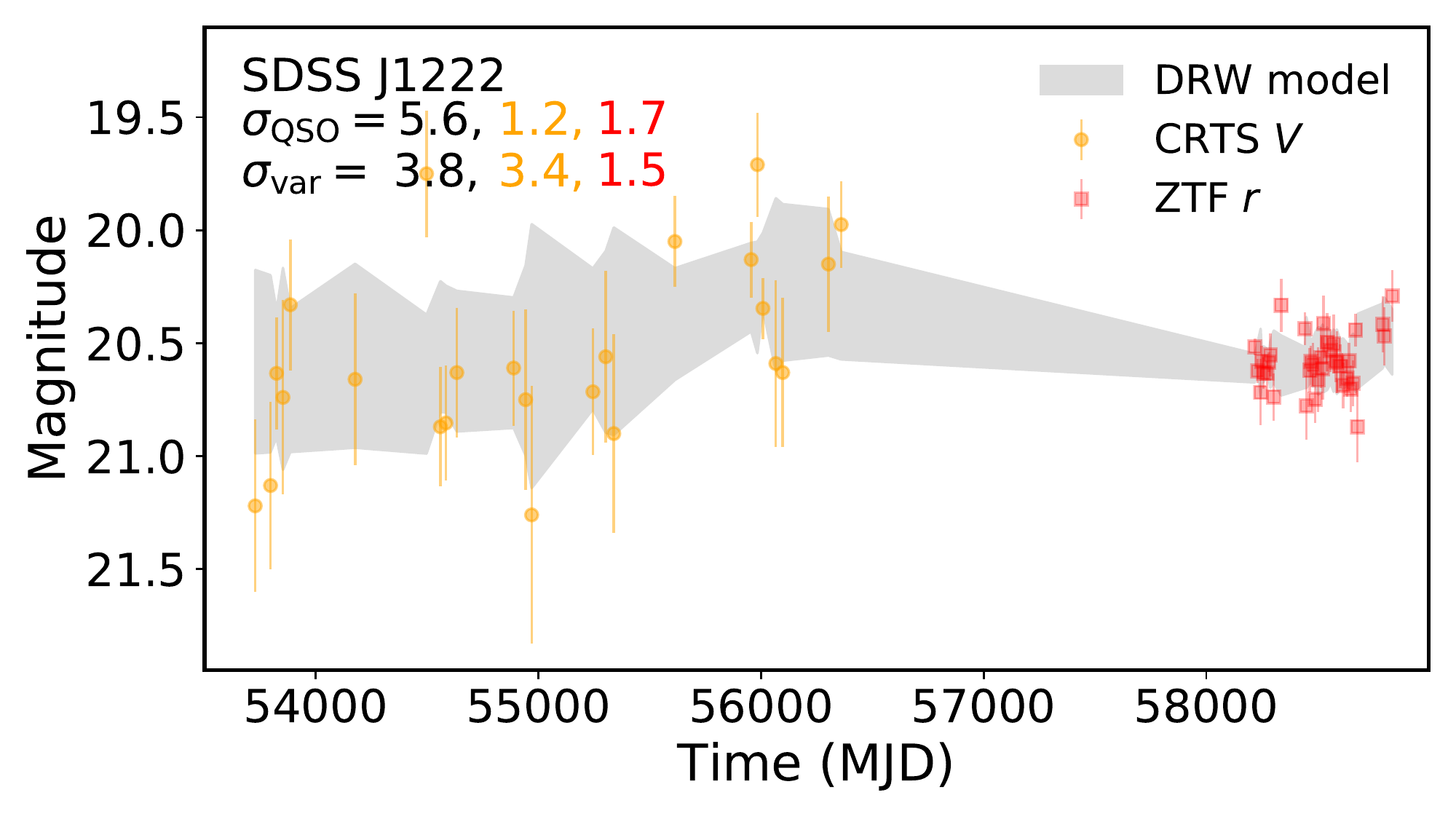}
    \caption{Light curves of the IT08 sample collected from CRTS \emph{V}-band and ZTF \emph{r}-band. The light curves are offset to match the median value of the CRTS data and binned in a window of 30 days (CRTS) or 10 days (ZTF). The fitted DRW model (${\pm}{1}{\sigma}$) is shown in light grey. The variability significance ($\sigma_{\text{var}}$) and significance the variability is DRW-like are shown ($\sigma_{\text{QSO}}$). The black text shows the combined significance, followed by the CRTS significance (orange) and ZTF significance (red). All galaxies with the possible exception of SDSS J0045 appear to exhibit low-levels of AGN-like variability. However, we cannot strongly rule-out the possibility that the variability in some of the sources is simply due to noise.}
    \label{fig:lc}
\end{figure*}

We queried light curves for the IT08 galaxies from the Catalina Real-Time Transient Survey (CRTS; \citealt{Drake2009}) and the Zwicky Transient Facility (ZTF; \citealt{Masci2019}). We perform outlier rejection in a 150 day moving window for the CRTS data and a 50 day moving window for the ZTF data. We offset the ZTF $r$-band data to match the median level of the CRTS $V$-band data. Then, we use the \textsc{qso\_fit} code\footnote{\url{http://butler.lab.asu.edu/qso_selection/index.html}} to estimate the variability significance ($\sigma_{\text{var}}$) and the significance the variability is well-described by the damped random walk (DRW)-like parameterisation of \citet{Butler2011} ($\sigma_{\text{QSO}}$). However, we caution that the limited quality of the light curves prevents tight constraints on a DRW model. In addition, the gap between the CRTS and ZTF data means our combined light curves could exclude real changes in the AGN luminosity. Therefore, we also fit the CRTS and ZTF data individually.

\subsection{VLA Imaging}

We downloaded archival VLA data for the four IT08 galaxies (VLA/10B-156; PI Henkel) and reduced and calibrated the continuum observations using the standard VLA reduction pipeline and \textsc{CASA} version 5.6.2. The sources were observed in 2010 in the C-band (5 GHz) in the C-configuration with a bandwidth of 256 MHz. We find no detections for any of the IT08 sources. We estimate the 5 GHz radio luminosity upper-limits (3$\sigma$) $L_{5~\text{GHz}}$ for each source as given in Table~\ref{tab:prop} using the standard equation:

\begin{equation}
L_\nu = \frac{S_{\rm obs} 4 \pi D_L^2}{(1+z)^{1+\alpha}}
\end{equation}

assuming a flat radio spectral index with $\alpha=0$ and $L_{5~\text{GHz}} = \nu L_{\nu}$ with $\nu=5$ GHz, where $D_L$ is the luminosity distance and $S_{\rm obs}$ is the observed flux density.

\section{Results}
\label{sec:results}

\begin{table*}
	\centering
	\caption{Observational properties of the IT08 sample. Values for the broad H$\alpha$ luminosity $L_{{\rm{H}}\alpha}$ and black hole mass $M_{{\bullet}}$ with statistical uncertainties are from our Gemini fitting, except for SDSS J0045, which we re-fit the SDSS spectrum. We re-compute each $M_{{\bullet}}$ using the relation of \citet{Reines2013}, noting that systematic uncertainties using the single-epoch virial method are typically $\sim$0.4 dex (e.g., \citealt{Shen2013}). The host galaxy stellar mass $M_{{\ast}}$ are estimated from the broadband SED in Appendix~\ref{sec:sed}. The Oxygen abundance $12 + \log \rm{O}/\rm{H}$, and X-ray luminosity $L_{2-10\ \rm{keV}}$ are from \citet{Simmonds2016}.}
	\label{tab:prop}
	\tiny
	\begin{tabular}{lccccccccc} 
		\hline
		Name & $z$ & $r$ & $L_{{\rm{H}}\alpha}$ & $\rm{FWHM}_{{\rm{H}}\alpha}$ & $ M_{{\bullet}}$ & $\log\ M_{\ast}$ & $12 + \log \rm{O}/\rm{H}$ & $L_{2-10\ \rm{keV}}$ & $L_{5\ \rm{GHz}}$\\
		& & (mag) & (erg s$^{-1}$) & (km s$^{-1}$) & ($M_{\odot}$) & ($\log\ M_{\odot}$) & (dex) & (erg s$^{-1}$) & (erg s$^{-1}$) \\
		\hline
		SDSS J004529.14+133908.6 & 0.29522 & 20.3 & $(4.5\pm0.6)\times10^{41}$ & $1465 \pm 369$ & $(5.6\pm2.9)\times10^6$ & $9.1\pm0.5$ & 7.9 & $<5.3\times10^{41}$ & $<2.3\times10^{38}$\\
		SDSS J102530.29+140207.3 & 0.10067 & 19.3 & $(4.4\pm0.1)\times10^{41}$ & $1962 \pm 45$ & $(1.0\pm0.1)\times10^7$ & $9.9\pm0.1$ & 7.4 & $<1.1\times10^{41}$ & $<2.6\times10^{37}$ \\
		SDSS J104755.92+073951.2 & 0.16828 & 18.8 & $(1.4\pm0.2)\times10^{42}$ & $1516 \pm 105$  &  $(1.0\pm0.2)\times10^7$& $10.0\pm0.1$ & 8.0 & $2.2\times10^{41}$ & $<7.3\times10^{37}$ \\
		SDSS J122245.71+360218.3 & 0.30112 & 20.0 & $(6.4\pm0.3)\times10^{41}$ & $1449 \pm 155$ & $(6.5\pm1.4)\times10^6$ & $8.4\pm0.4$ & 7.9 & $<5.5\times10^{41}$ & $<2.5\times10^{38}$ \\
		\hline
	\end{tabular}
\end{table*}

The data and spectral fitting of the H$\alpha$-[N~\textsc{ii}] complex for each source is shown in Fig.~\ref{fig:spectra} along with the SDSS spectral epochs (modeled in the same manner) for comparison. We plot the broad H$\alpha$ luminosities of the IT08 sample versus time including our new Gemini data and the values from \citet{Simmonds2016} in Fig.~\ref{fig:Halphaluminosity}. For comparison, we plot the transients PHL 293B \citep{Burke2020b} and SDSS1133 \citep{Koss2014}, and the luminous long-lived SNe 1988Z \citep{Aretxaga1999,Smith2017} and 2005ip \citep{smith2009,Smith2017}.

We find the broad H$\alpha$ emission is still persistent in 2020 at roughly the same levels as before to the $\sim$20 per cent level. Decomposing the narrow and broad-line components can sometimes be ambiguous, therefore we caution on apparent differences in the broad-line measurements between the SDSS and Gemini epochs. The large uncertainties in the broad-line width in the SDSS epoch of J1222 reflect the ambiguity of such decomposition. Given the additional systematic uncertainties due to aperture effects, instrumental resolution, and photometric conditions, we find no evidence for strong variability of the broad Balmer emission in the IT08 sample. We use the broad H$\alpha$ luminosities and FWHM in Fig.~\ref{fig:spectra} to compute virial black hole masses of $M_{\bullet} \approx 10^{6.7}{-10^{7.0}\ M_{\odot}}$ using the updated relation of \citet{Reines2013}:

\begin{equation}
\begin{split}
    \log{\left(\frac{M_{\bullet} }{M_{\odot}} \right)} = 6.57 + 0.47 \log{ \left( \frac{L_{\rm{H}\alpha}}{10^{42}\ \rm{ erg\ s}^{-1}} \right) } \\ + 2.06 \log{ \left( \frac{\rm{FWHM}_{\rm{H}\alpha}}{10^{3}\ \rm{ km\ s}^{-1}} \right) }
\end{split}
\end{equation}

where $L_{\rm{H}\alpha}$ and $\rm{FWHM}_{\rm{H}\alpha}$ are the broad H$\alpha$ luminosity and FWHM.

We show the extracted light curves in Fig.~\ref{fig:lc}. Low-levels of variability are detected in all IT08 galaxies with the possible exception of SDSS J0045. The variability in these three galaxies show red noise, as expected in the commonly-used stochastic DRW model of AGN variability \citep{Kelly2009,MacLeod2010}. This detection of variability in these three galaxies supports the AGN scenario and likely indicates the AGNs are not strongly obscured.

A summary of the observational properties of the IT08 sample is given in Table~\ref{tab:prop}. The AGNs lie 1--2 dex below the expected X-ray relations of \citet{Panessa2006}. Using the X-ray luminosity upper-limits and black hole masses in Table~\ref{tab:prop}, we expect 5 GHz radio luminosities of less than 37.4 -- 37.7 erg s$^{-1}$ using the ``fundamental plane'' relation of \citet{Merloni2003}. Therefore, the lack of radio detections is consistent with the IT08 sample being intrinsically X-ray weak.

\subsection{SDSS J1025: Balmer Absorption Line}

We identify a $119\pm46$ km s$^{-1}$ blueshifted (relative to the broad H$\alpha$) narrow absorption line in the spectrum of SDSS J1025. This line is real, and not due to telluric absorption. It is present at lower spectral resolution in the SDSS spectrum at the same blueshift. We checked other spectra on the same SDSS plate (observed at the same time) with similar flux levels near 7220\ \AA\ and found no similar features. This feature apparently went unnoticed until now, likely due to insufficient spectral resolution and binning/smoothing of the data obscuring the line. 

Balmer H$\alpha$ absorption in AGNs is very unusual, with only a handful reported in the literature. This is usually interpreted as an indicator of an AGN-driven outflow with high hydrogen column densities (e.g. \citealt{Hutchings2002,Aoki2010,Wang2015}). These absorption features bear some resemblance to P Cygni profiles, similar to what was seen in the galaxy PHL 293B. However, unlike PHL 293B, no additional absorption features are obvious in the spectrum of SDSS J1025, the absorption line is narrower, and luminosities much larger.

Following \citet{Wang2015}, we estimate the neutral hydrogen column density in the optically thin regime using the equation,


\begin{equation}
    N_{\text{HI,2}} = 1.130\times10^{12} \text{ cm}^{-1} \frac{EW_{\lambda} }{ f \lambda^2 }
\label{eq:EW}
\end{equation}

where $EW_\lambda$ is the equivalent-width of the absorption line, $f$ is the oscillator strength. Using the $EW_\lambda$ of the H$\alpha$ absorption of 39.3\ \AA\ from our modeling of the Gemini spectrum and $f=0.64$, we find a neutral hydrogen column density in the $n=2$ shell of $N_{\text{HI,2}} = 1.6\times10^{14}$ cm$^{-2}$. We can estimate the $n=1$ level population produced by Ly$\alpha$ trapping in thermal equilibrium using the equation given by \citet{Hall2007},

\begin{equation}
    \frac{N_{\rm{HI,1}}}{N_{\rm{HI,2}}} = \frac{1}{4\tau_{\text{Ly}\alpha}} e^{10.2\text{ eV}/kT}
\label{eq:NINIIfromtauLya}
\end{equation}

where the optical depth at the center of the Ly$\alpha$ absorption is $\tau_{\text{Ly}\alpha}=0.12\ \tau_{\text{H}\alpha}\ {N_{\text{HI,1}}}/{N_{\text{HI,2}}}$ \citep{Aoki2010,Wang2015}. Substituting this into Eq.~\ref{eq:NINIIfromtauLya}, we obtain,

\begin{equation}
    \frac{N_{\rm{HI,1}}}{N_{\rm{HI,2}}} = \frac{1.44}{\sqrt{\tau_{\text{H}\alpha}}} e^{5.1\text{ eV}/kT}.
\label{eq:NINIIfromtauHa}
\end{equation}

To estimate the optical depth of H$\alpha$, we use $\tau_{\text{H}\alpha}=1.69\times10^{5}\ EW_\lambda b / \lambda^2$ where the Doppler broadening parameter $b \approx \text{FWHM}/1.665 = 83.9$ \AA\ measured from our modeling of the Gemini absorption line. We find $\tau_{\text{H}\alpha}=12.9$. Taking $T=7500$ K \citep{Osterbrock2006}, we find $N_{\text{HI,1}} = 1.6\times10^{17}$ cm$^{-2}$ using Eq.~\ref{eq:NINIIfromtauHa}. Finally we find a total column density of $N_{\text{HI}} \approx N_{\text{HI,1}} + N_{\text{HI,2}}= 1.6\times10^{17}$ cm$^{-2}$.





Comparing $N_{\text{HI}}$ to the cross section due to Thomson scattering $\sigma_{\text{T}} \approx 10^{-24}$ cm$^{2}$, this implies the AGN is not Compton thick. Therefore the AGN is unobscured and the observed X-ray luminosity should be close to the true value. \citet{Simmonds2016} estimate the upper-limit of the hard X-ray luminosity to be $L_{2-10\text{ keV}} < 1.1\times10^{41}$ erg s$^{-1}$ for SDSS J1025 assuming a power-law spectrum of $\Gamma=1.8$. We quote their upper-limits for all IT08 AGNs in Table~\ref{tab:prop}.

\section{Prospects for Future surveys}
\label{sec:prospects}

\begin{figure}
	\includegraphics[width=0.5\textwidth]{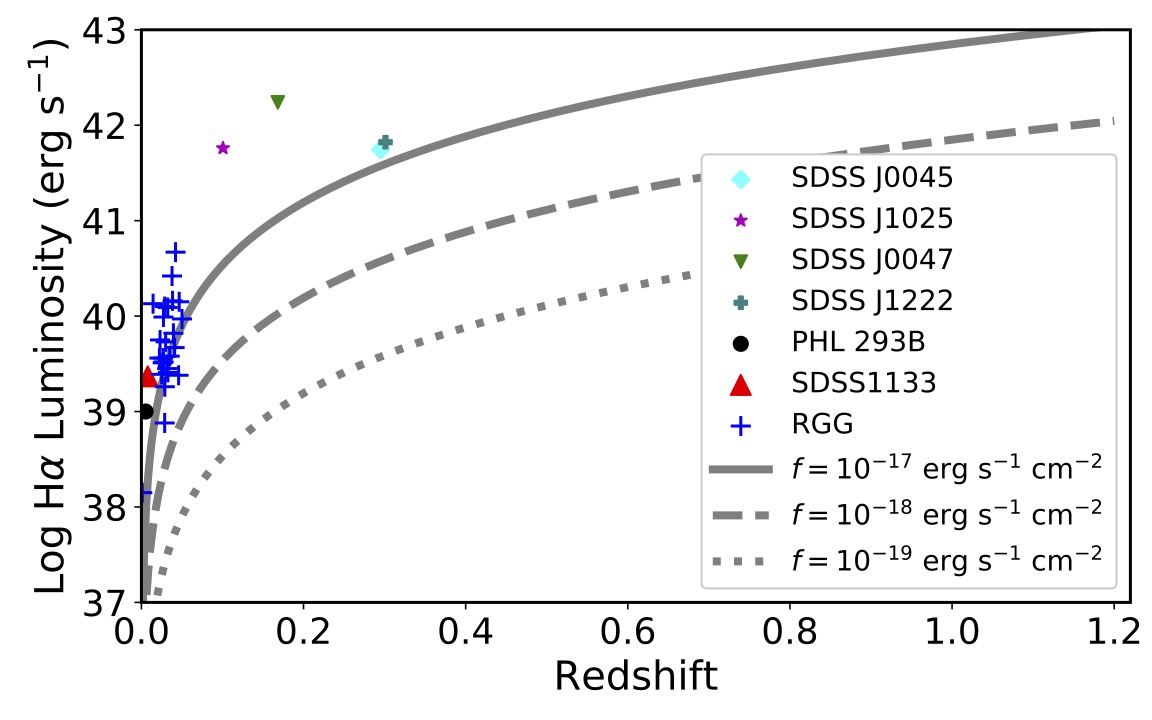}
    \caption{Broad H$\alpha$ luminosity versus redshift for the IT08 sample (diamonds, stars, inverted triangles, plus symbols) compared to PHL 293B (black circles), SDSS1133 (red triangles), and the sample of \citet{Reines2013} broad-line AGNs (RGG; blue $+$ symbols). Varying sensitivity curves are shown as grey lines assuming a broad H$\alpha$ line typical of the IT08 AGNs.}
    \label{fig:sensitivity}
\end{figure}

The small but non-zero sample size of metal-poor AGNs \citep{Izotov2008,Cann2020,Reines2013,Reines2014} motivates future searches with current and upcoming facilities. To investigate this, we plot the broad H$\alpha$ luminosity sensitivity curve versus redshift assuming a Gaussian H$\alpha$ broad emission line with $\text{FHWM} = 2000$ km s$^{-1}$ and luminosity $5\times10^{41}$ erg s$^{-1}$ at varying spectral-line flux sensitivities in Fig.~\ref{fig:sensitivity}. If we require the H$\alpha$ line to be detected, this limits us to redshifts $z<0.5$ assuming a spectral coverage $< 9800$ \AA. This demonstrates that very deep near-infrared spectra, likely using 30-m class telescopes will be necessary to detect similar AGNs at $z\sim1$.

New detections at low-$z$ are more promising. The detection rate of the IT08 sample is 4/675,000 or 0.0006 per cent of the full SDSS DR4 sample of galaxies \cite{Izotov2008}. The Dark Energy Spectroscopic Experiment (DESI) will conduct a flux-limited survey of $10^7$ galaxies at a median redshift of 0.2 \citep{Levi2019}. Roughly half of the DESI targets will be ELGs \citep{Raichoor2020}. Assuming a flux sensitivity similar to SDSS, we can expect to find about 60 additional IT08-like AGNs. Other upcoming surveys, such as the Subaru Prime Focus Spectrograph Survey \citep{Takada2014} are also promising.

The main problem is having two spectral epochs at least 10 years apart with sufficient $S/N$ to identify broad Balmer emission lines. Combined with optical photometric data, one could conceivably identify AGNs in metal-poor galaxies with high confidence without multiple spectral epochs. \citet{Baldassare2018} demonstrated the capability of SDSS imaging to identify dwarf AGNs using long-term variability. \citet{Guo2020} demonstrated the capability of Dark Energy Survey photometry to identify dwarf AGNs at intermediate redshifts of $z\sim0.8$. A dwarf Type I AGN similar to the archetypal example NGC 4395 with SMBH mass of $10^5 M_{\odot}$ will vary photometrically at the $\sim$1--10 per cent level on a rest-frame timescale of a few days \citep{Kelly2009,Burke2020a}. The variability is stochastic, therefore can be distinguished from stellar transients (e.g. SN flares and tidal disruption events) which generally fade on weeks to years timescales. Our light curve analysis in this work demonstrates the capability of photometric surveys to aid in AGN identification when it would otherwise be ambiguous.

\section{Conclusions}
\label{sec:conclusions}

The H$\alpha$ luminosities of the IT08 galaxies are orders of magnitude larger than what is seen in stellar transients in metal-poor galaxies. Their broad H$\alpha$ duration of $\sim$15 years or longer leaves us with the AGN interpretation. This is further supported by the low-level photometric variability similar to what is seen in Type I Seyferts. The AGNs appear to be intrinsically X-ray and radio weak, lying 1--2 dex below the relations of \citet{Panessa2006}.

We identify SDSS J1025 as a new Balmer-absorption AGN. According to \citet{Wang2015}, only eight are reported in the literature (including their AGN). We interpret this as evidence for an AGN-driven outflow. This raises the interesting possibility that the metals in these galaxies may be blown-out by an AGN-driven wind. The velocity offset of the line of $\sim$119 km s$^{-1}$ indicates an extremely slow wind compared to what is typically considered for broad absorption line quasars \citep{Faucher-Guiguere2012}. This may perhaps be understood given the very extreme physical regime (low metallicity, much smaller black hole masses) of the IT08 AGNs. The column density of neutral hydrogen derived from the absorption line indicates the they are not strongly obscured in the X-ray. This probably indicates the IT08 AGNs are intrinsically X-ray weak.


Recently, \citet{Cann2020} identified a low-metallicity AGN with an X-ray luminosity lying $\sim$2 dex below the prediction from the mid-infrared luminosity relation. Like the IT08 objects, it also lies near the boundary between the star-forming and AGN regions of the BPT diagram. This suggests a new population of X-ray/radio-weak low-metallicity dwarf AGNs in star-forming galaxies, perhaps with unusually low accretion rates. This motivates more sensitive X-ray observations of larger samples of low-metallicity, low-mass AGNs.

Nevertheless, the connection between metallicity and X-ray emission in AGNs is unclear at present. \citet{Simmonds2016} suggest the IT08 AGNs may fail to generate a hard X-ray emitting corona. Perhaps metal-poor AGNs have soft X-ray spectra, a phenomena seen in some broad-line dwarf AGNs, such as Pox 52 \citep{Barth2004,Dewangan2008}. Another possibility is the black hole masses of the IT08 AGNs are significantly over-estimated by $\sim 1$ dex in $M_{\odot}$ using single-epoch virial relations, perhaps due to a different broad-line region geometry. However, there is simply insufficient understanding of AGNs in metal-poor star-forming environments to venture beyond speculation.

The low X-ray luminosities and classifications as star-forming galaxies using BPT diagnostics underscores the difficulty of identifying metal-deficient AGNs using traditional methods. Future studies investigating the AGN-host galaxy scaling relations in metal-poor AGNs are a promising avenue toward a more complete understanding of SMBH growth and formation.

\section*{Acknowledgements}

C.J.B. and Y.C.C. acknowledge the Illinois Graduate Survey Science Fellowship for support. Y.S. acknowledges support from an Alfred P. Sloan Research Fellowship and NSF grant AST-2009947. H.G. acknowledges support from NSF grant AST-1907290. We thank Kedar Phadke and Kaiwen Zhang for their help with using SED fitting codes. We thank the anonymous referee for constructive comments which improved this work.

Based on observations obtained at the Gemini Observatory, which is operated by the Association of Universities for Research in Astronomy, Inc., under a cooperative agreement with the NSF on behalf of the Gemini partnership: the National Science Foundation (United States), National Research Council (Canada), CONICYT (Chile), Ministerio de Ciencia, Tecnolog\'{i}a e Innovaci\'{o}n Productiva (Argentina), Minist\'{e}rio da Ci\^{e}ncia, Tecnologia e Inova\c{c}\~{a}o (Brazil), and Korea Astronomy and Space Science Institute (Republic of Korea).

The National Radio Astronomy Observatory is a facility of the National Science Foundation operated under cooperative agreement by Associated Universities, Inc.

The CSS survey is funded by the National Aeronautics and Space
Administration under Grant No. NNG05GF22G issued through the Science
Mission Directorate Near-Earth Objects Observations Program.  The CRTS
survey is supported by the U.S.~National Science Foundation under
grants AST-0909182 and AST-1313422.


Based on observations obtained with the Samuel Oschin 48-inch Telescope at the Palomar Observatory as part of the Zwicky Transient Facility project. ZTF is supported by the National Science Foundation under Grant No. AST-1440341 and a collaboration including Caltech, IPAC, the Weizmann Institute for Science, the Oskar Klein Center at Stockholm University, the University of Maryland, the University of Washington, Deutsches Elektronen-Synchrotron and Humboldt University, Los Alamos National Laboratories, the TANGO Consortium of Taiwan, the University of Wisconsin at Milwaukee, and Lawrence Berkeley National Laboratories. Operations are conducted by COO, IPAC, and UW.

\section*{Data Availability}

Gemini spectra will be available on the Gemini Observatory Archive at \url{https://archive.gemini.edu/searchform} after the proprietary period. SDSS spectra are available at \url{https://www.sdss.org/dr16}. The VLA data are available at the VLA data archive \url{https://science.nrao.edu/facilities/vla/archive/index}. CRTS light curves are available at \url{http://nesssi.cacr.caltech.edu/DataRelease}. ZTF light curves can be downloaded at \url{https://irsa.ipac.caltech.edu/Missions/ztf.html}. The broadband photometry are queried from the VizieR photometry tool at \url{http://vizier.unistra.fr/vizier/sed/}.



\bibliographystyle{mnras}
\bibliography{example} 

\begin{thebibliography}{}
\makeatletter
\relax
\def\mn@urlcharsother{\let\do\@makeother \do\$\do\&\do\#\do\^\do\_\do\%\do\~}
\def\mn@doi{\begingroup\mn@urlcharsother \@ifnextchar [ {\mn@doi@}
  {\mn@doi@[]}}
\def\mn@doi@[#1]#2{\def\@tempa{#1}\ifx\@tempa\@empty \href
  {http://dx.doi.org/#2} {doi:#2}\else \href {http://dx.doi.org/#2} {#1}\fi
  \endgroup}
\def\mn@eprint#1#2{\mn@eprint@#1:#2::\@nil}
\def\mn@eprint@arXiv#1{\href {http://arxiv.org/abs/#1} {{\tt arXiv:#1}}}
\def\mn@eprint@dblp#1{\href {http://dblp.uni-trier.de/rec/bibtex/#1.xml}
  {dblp:#1}}
\def\mn@eprint@#1:#2:#3:#4\@nil{\def\@tempa {#1}\def\@tempb {#2}\def\@tempc
  {#3}\ifx \@tempc \@empty \let \@tempc \@tempb \let \@tempb \@tempa \fi \ifx
  \@tempb \@empty \def\@tempb {arXiv}\fi \@ifundefined
  {mn@eprint@\@tempb}{\@tempb:\@tempc}{\expandafter \expandafter \csname
  mn@eprint@\@tempb\endcsname \expandafter{\@tempc}}}

\bibitem[\protect\citeauthoryear{{Allan}, {Groh}, {Mehner}, {Smith}, {Boian},
  {Farrell}  \& {Andrews}}{{Allan} et~al.}{2020}]{Allan2020}
{Allan} A.~P.,  {Groh} J.~H.,  {Mehner} A.,  {Smith} N.,  {Boian} I.,
  {Farrell} E.~J.,   {Andrews} J.~E.,  2020, \mn@doi [\mnras]
  {10.1093/mnras/staa1629}, \href
  {https://ui.adsabs.harvard.edu/abs/2020MNRAS.496.1902A} {496, 1902}

\bibitem[\protect\citeauthoryear{{Aoki}}{{Aoki}}{2010}]{Aoki2010}
{Aoki} K.,  2010, \mn@doi [\pasj] {10.1093/pasj/62.5.1333}, \href
  {https://ui.adsabs.harvard.edu/abs/2010PASJ...62.1333A} {62, 1333}

\bibitem[\protect\citeauthoryear{{Aretxaga}, {Benetti}, {Terlevich}, {Fabian},
  {Cappellaro}, {Turatto}  \& {della Valle}}{{Aretxaga}
  et~al.}{1999}]{Aretxaga1999}
{Aretxaga} I.,  {Benetti} S.,  {Terlevich} R.~J.,  {Fabian} A.~C.,
  {Cappellaro} E.,  {Turatto} M.,   {della Valle} M.,  1999, \mn@doi [\mnras]
  {10.1046/j.1365-8711.1999.02830.x}, \href
  {https://ui.adsabs.harvard.edu/abs/1999MNRAS.309..343A} {309, 343}

\bibitem[\protect\citeauthoryear{{Baldassare} et~al.,}{{Baldassare}
  et~al.}{2016}]{Baldassare2016}
{Baldassare} V.~F.,  et~al., 2016, \mn@doi [\apj] {10.3847/0004-637X/829/1/57},
  \href {https://ui.adsabs.harvard.edu/abs/2016ApJ...829...57B} {829, 57}

\bibitem[\protect\citeauthoryear{{Baldassare}, {Geha}  \&
  {Greene}}{{Baldassare} et~al.}{2018}]{Baldassare2018}
{Baldassare} V.~F.,  {Geha} M.,   {Greene} J.,  2018, \mn@doi [\apj]
  {10.3847/1538-4357/aae6cf}, \href
  {https://ui.adsabs.harvard.edu/abs/2018ApJ...868..152B} {868, 152}

\bibitem[\protect\citeauthoryear{{Barth}, {Ho}, {Rutledge}  \&
  {Sargent}}{{Barth} et~al.}{2004}]{Barth2004}
{Barth} A.~J.,  {Ho} L.~C.,  {Rutledge} R.~E.,   {Sargent} W. L.~W.,  2004,
  \mn@doi [\apj] {10.1086/383302}, \href
  {https://ui.adsabs.harvard.edu/abs/2004ApJ...607...90B} {607, 90}

\bibitem[\protect\citeauthoryear{{Boquien}, {Burgarella}, {Roehlly}, {Buat},
  {Ciesla}, {Corre}, {Inoue}  \& {Salas}}{{Boquien} et~al.}{2019}]{Boquien2019}
{Boquien} M.,  {Burgarella} D.,  {Roehlly} Y.,  {Buat} V.,  {Ciesla} L.,
  {Corre} D.,  {Inoue} A.~K.,   {Salas} H.,  2019, \mn@doi [\aap]
  {10.1051/0004-6361/201834156}, \href
  {https://ui.adsabs.harvard.edu/abs/2019A&A...622A.103B} {622, A103}

\bibitem[\protect\citeauthoryear{{Bruzual} \& {Charlot}}{{Bruzual} \&
  {Charlot}}{2003}]{Bruzual2003}
{Bruzual} G.,  {Charlot} S.,  2003, \mn@doi [\mnras]
  {10.1046/j.1365-8711.2003.06897.x}, \href
  {https://ui.adsabs.harvard.edu/abs/2003MNRAS.344.1000B} {344, 1000}

\bibitem[\protect\citeauthoryear{{Burgarella}, {Buat}  \&
  {Iglesias-P{\'a}ramo}}{{Burgarella} et~al.}{2005}]{Burgarella2005}
{Burgarella} D.,  {Buat} V.,   {Iglesias-P{\'a}ramo} J.,  2005, \mn@doi
  [\mnras] {10.1111/j.1365-2966.2005.09131.x}, \href
  {https://ui.adsabs.harvard.edu/abs/2005MNRAS.360.1413B} {360, 1413}

\bibitem[\protect\citeauthoryear{{Burke} et~al.,}{{Burke}
  et~al.}{2020a}]{Burke2020b}
{Burke} C.~J.,  et~al., 2020a, \mn@doi [\apjl] {10.3847/2041-8213/ab88de},
  \href {https://ui.adsabs.harvard.edu/abs/2020ApJ...894L...5B} {894, L5}

\bibitem[\protect\citeauthoryear{{Burke}, {Shen}, {Chen}, {Scaringi},
  {Faucher-Giguere}, {Liu}  \& {Yang}}{{Burke} et~al.}{2020b}]{Burke2020a}
{Burke} C.~J.,  {Shen} Y.,  {Chen} Y.-C.,  {Scaringi} S.,  {Faucher-Giguere}
  C.-A.,  {Liu} X.,   {Yang} Q.,  2020b, \mn@doi [\apj]
  {10.3847/1538-4357/aba3ce}, \href
  {https://ui.adsabs.harvard.edu/abs/2020ApJ...899..136B} {899, 136}

\bibitem[\protect\citeauthoryear{{Butler} \& {Bloom}}{{Butler} \&
  {Bloom}}{2011}]{Butler2011}
{Butler} N.~R.,  {Bloom} J.~S.,  2011, \mn@doi [\aj]
  {10.1088/0004-6256/141/3/93}, \href
  {https://ui.adsabs.harvard.edu/abs/2011AJ....141...93B} {141, 93}

\bibitem[\protect\citeauthoryear{{Calzetti}, {Armus}, {Bohlin}, {Kinney},
  {Koornneef}  \& {Storchi-Bergmann}}{{Calzetti} et~al.}{2000}]{Calzetti2000}
{Calzetti} D.,  {Armus} L.,  {Bohlin} R.~C.,  {Kinney} A.~L.,  {Koornneef} J.,
   {Storchi-Bergmann} T.,  2000, \mn@doi [\apj] {10.1086/308692}, \href
  {https://ui.adsabs.harvard.edu/abs/2000ApJ...533..682C} {533, 682}

\bibitem[\protect\citeauthoryear{{Cann} et~al.,}{{Cann}
  et~al.}{2020}]{Cann2020}
{Cann} J.~M.,  et~al., 2020, \mn@doi [\apj] {10.3847/1538-4357/ab8b64}, \href
  {https://ui.adsabs.harvard.edu/abs/2020ApJ...895..147C} {895, 147}

\bibitem[\protect\citeauthoryear{{Cardamone} et~al.,}{{Cardamone}
  et~al.}{2009}]{Cardamone2009}
{Cardamone} C.,  et~al., 2009, \mn@doi [\mnras]
  {10.1111/j.1365-2966.2009.15383.x}, \href
  {https://ui.adsabs.harvard.edu/abs/2009MNRAS.399.1191C} {399, 1191}

\bibitem[\protect\citeauthoryear{{Chabrier}}{{Chabrier}}{2003}]{Chabrier2003}
{Chabrier} G.,  2003, \mn@doi [\pasp] {10.1086/376392}, \href
  {https://ui.adsabs.harvard.edu/abs/2003PASP..115..763C} {115, 763}

\bibitem[\protect\citeauthoryear{{Ciesla} et~al.,}{{Ciesla}
  et~al.}{2015}]{Ciesla2015}
{Ciesla} L.,  et~al., 2015, \mn@doi [\aap] {10.1051/0004-6361/201425252}, \href
  {https://ui.adsabs.harvard.edu/abs/2015A&A...576A..10C} {576, A10}

\bibitem[\protect\citeauthoryear{{Dewangan}, {Mathur}, {Griffiths}  \&
  {Rao}}{{Dewangan} et~al.}{2008}]{Dewangan2008}
{Dewangan} G.~C.,  {Mathur} S.,  {Griffiths} R.~E.,   {Rao} A.~R.,  2008,
  \mn@doi [\apj] {10.1086/591728}, \href
  {https://ui.adsabs.harvard.edu/abs/2008ApJ...689..762D} {689, 762}

\bibitem[\protect\citeauthoryear{{Draine} et~al.,}{{Draine}
  et~al.}{2007}]{Draine2007}
{Draine} B.~T.,  et~al., 2007, \mn@doi [\apj] {10.1086/518306}, \href
  {https://ui.adsabs.harvard.edu/abs/2007ApJ...663..866D} {663, 866}

\bibitem[\protect\citeauthoryear{{Draine} et~al.,}{{Draine}
  et~al.}{2014}]{Draine2014}
{Draine} B.~T.,  et~al., 2014, \mn@doi [\apj] {10.1088/0004-637X/780/2/172},
  \href {https://ui.adsabs.harvard.edu/abs/2014ApJ...780..172D} {780, 172}

\bibitem[\protect\citeauthoryear{{Drake} et~al.,}{{Drake}
  et~al.}{2009}]{Drake2009}
{Drake} A.~J.,  et~al., 2009, \mn@doi [\apj] {10.1088/0004-637X/696/1/870},
  \href {https://ui.adsabs.harvard.edu/abs/2009ApJ...696..870D} {696, 870}

\bibitem[\protect\citeauthoryear{{Faucher-Gigu{\`e}re} \&
  {Quataert}}{{Faucher-Gigu{\`e}re} \& {Quataert}}{2012}]{Faucher-Guiguere2012}
{Faucher-Gigu{\`e}re} C.-A.,  {Quataert} E.,  2012, \mn@doi [\mnras]
  {10.1111/j.1365-2966.2012.21512.x}, \href
  {https://ui.adsabs.harvard.edu/abs/2012MNRAS.425..605F} {425, 605}

\bibitem[\protect\citeauthoryear{{Greene}}{{Greene}}{2012}]{Greene2012}
{Greene} J.~E.,  2012, \mn@doi [Nature Communications] {10.1038/ncomms2314},
  \href {https://ui.adsabs.harvard.edu/abs/2012NatCo...3.1304G} {3, 1304}

\bibitem[\protect\citeauthoryear{Groves, Heckman  \& Kauffmann}{Groves
  et~al.}{2006}]{Groves2006}
Groves B.~A.,  Heckman T.~M.,   Kauffmann G.,  2006, \mn@doi [MNRAS]
  {10.1111/j.1365-2966.2006.10812.x}, 371, 1559

\bibitem[\protect\citeauthoryear{{Guo}, {Shen}  \& {Wang}}{{Guo}
  et~al.}{2018}]{Guo2018}
{Guo} H.,  {Shen} Y.,   {Wang} S.,  2018, {PyQSOFit: Python code to fit the
  spectrum of quasars} (\mn@eprint {ascl} {1809.008})

\bibitem[\protect\citeauthoryear{{Guo} et~al.,}{{Guo} et~al.}{2020}]{Guo2020}
{Guo} H.,  et~al., 2020, \mn@doi [\mnras] {10.1093/mnras/staa1803}, \href
  {https://ui.adsabs.harvard.edu/abs/2020MNRAS.496.3636G} {496, 3636}

\bibitem[\protect\citeauthoryear{{Hall}}{{Hall}}{2007}]{Hall2007}
{Hall} P.~B.,  2007, \mn@doi [\aj] {10.1086/511272}, \href
  {https://ui.adsabs.harvard.edu/abs/2007AJ....133.1271H} {133, 1271}

\bibitem[\protect\citeauthoryear{{Hutchings}, {Crenshaw}, {Kraemer}, {Gabel},
  {Kaiser}, {Weistrop}  \& {Gull}}{{Hutchings} et~al.}{2002}]{Hutchings2002}
{Hutchings} J.~B.,  {Crenshaw} D.~M.,  {Kraemer} S.~B.,  {Gabel} J.~R.,
  {Kaiser} M.~E.,  {Weistrop} D.,   {Gull} T.~R.,  2002, \mn@doi [\aj]
  {10.1086/344080}, \href
  {https://ui.adsabs.harvard.edu/abs/2002AJ....124.2543H} {124, 2543}

\bibitem[\protect\citeauthoryear{{Inoue}}{{Inoue}}{2011}]{Inoue2011}
{Inoue} A.~K.,  2011, \mn@doi [\mnras] {10.1111/j.1365-2966.2011.18906.x},
  \href {https://ui.adsabs.harvard.edu/abs/2011MNRAS.415.2920I} {415, 2920}

\bibitem[\protect\citeauthoryear{{Izotov} \& {Thuan}}{{Izotov} \&
  {Thuan}}{2008}]{Izotov2008}
{Izotov} Y.~I.,  {Thuan} T.~X.,  2008, \mn@doi [\apj] {10.1086/591660}, \href
  {https://ui.adsabs.harvard.edu/abs/2008ApJ...687..133I} {687, 133}

\bibitem[\protect\citeauthoryear{{Izotov} \& {Thuan}}{{Izotov} \&
  {Thuan}}{2009}]{Izotov2009}
{Izotov} Y.~I.,  {Thuan} T.~X.,  2009, \mn@doi [\apj]
  {10.1088/0004-637X/690/2/1797}, \href
  {https://ui.adsabs.harvard.edu/abs/2009ApJ...690.1797I} {690, 1797}

\bibitem[\protect\citeauthoryear{{Izotov}, {Thuan}  \& {Guseva}}{{Izotov}
  et~al.}{2007}]{Izotov2007}
{Izotov} Y.~I.,  {Thuan} T.~X.,   {Guseva} N.~G.,  2007, \mn@doi [\apj]
  {10.1086/522923}, \href
  {https://ui.adsabs.harvard.edu/abs/2007ApJ...671.1297I} {671, 1297}

\bibitem[\protect\citeauthoryear{{Izotov}, {Guseva}, {Fricke}  \&
  {Henkel}}{{Izotov} et~al.}{2011}]{Izotov2011}
{Izotov} Y.~I.,  {Guseva} N.~G.,  {Fricke} K.~J.,   {Henkel} C.,  2011, \mn@doi
  [\aap] {10.1051/0004-6361/201016296}, \href
  {https://ui.adsabs.harvard.edu/abs/2011A&A...533A..25I} {533, A25}

\bibitem[\protect\citeauthoryear{{Kelly}, {Bechtold}  \&
  {Siemiginowska}}{{Kelly} et~al.}{2009}]{Kelly2009}
{Kelly} B.~C.,  {Bechtold} J.,   {Siemiginowska} A.,  2009, \mn@doi [\apj]
  {10.1088/0004-637X/698/1/895}, \href
  {https://ui.adsabs.harvard.edu/abs/2009ApJ...698..895K} {698, 895}

\bibitem[\protect\citeauthoryear{{Koss} et~al.,}{{Koss}
  et~al.}{2014}]{Koss2014}
{Koss} M.,  et~al., 2014, \mn@doi [\mnras] {10.1093/mnras/stu1673}, \href
  {https://ui.adsabs.harvard.edu/abs/2014MNRAS.445..515K} {445, 515}

\bibitem[\protect\citeauthoryear{{Lawrence} et~al.,}{{Lawrence}
  et~al.}{2007}]{Lawrence2007}
{Lawrence} A.,  et~al., 2007, \mn@doi [\mnras]
  {10.1111/j.1365-2966.2007.12040.x}, \href
  {https://ui.adsabs.harvard.edu/abs/2007MNRAS.379.1599L} {379, 1599}

\bibitem[\protect\citeauthoryear{{Leitherer}, {Li}, {Calzetti}  \&
  {Heckman}}{{Leitherer} et~al.}{2002}]{Leitherer2002}
{Leitherer} C.,  {Li} I.~H.,  {Calzetti} D.,   {Heckman} T.~M.,  2002, \mn@doi
  [\apjs] {10.1086/342486}, \href
  {https://ui.adsabs.harvard.edu/abs/2002ApJS..140..303L} {140, 303}

\bibitem[\protect\citeauthoryear{{Levi} et~al.,}{{Levi}
  et~al.}{2019}]{Levi2019}
{Levi} M.,  et~al., 2019, in Bulletin of the American Astronomical Society.
  p.~57 (\mn@eprint {arXiv} {1907.10688})

\bibitem[\protect\citeauthoryear{{MacLeod} et~al.,}{{MacLeod}
  et~al.}{2010}]{MacLeod2010}
{MacLeod} C.~L.,  et~al., 2010, \mn@doi [\apj] {10.1088/0004-637X/721/2/1014},
  \href {https://ui.adsabs.harvard.edu/abs/2010ApJ...721.1014M} {721, 1014}

\bibitem[\protect\citeauthoryear{{Martin} et~al.,}{{Martin}
  et~al.}{2005}]{Martin2005}
{Martin} D.~C.,  et~al., 2005, \mn@doi [\apjl] {10.1086/426387}, \href
  {https://ui.adsabs.harvard.edu/abs/2005ApJ...619L...1M} {619, L1}

\bibitem[\protect\citeauthoryear{{Masci} et~al.,}{{Masci}
  et~al.}{2019}]{Masci2019}
{Masci} F.~J.,  et~al., 2019, \mn@doi [\pasp] {10.1088/1538-3873/aae8ac}, \href
  {https://ui.adsabs.harvard.edu/abs/2019PASP..131a8003M} {131, 018003}

\bibitem[\protect\citeauthoryear{{Merloni}, {Heinz}  \& {di Matteo}}{{Merloni}
  et~al.}{2003}]{Merloni2003}
{Merloni} A.,  {Heinz} S.,   {di Matteo} T.,  2003, \mn@doi [\mnras]
  {10.1046/j.1365-2966.2003.07017.x}, \href
  {https://ui.adsabs.harvard.edu/abs/2003MNRAS.345.1057M} {345, 1057}

\bibitem[\protect\citeauthoryear{{Mezcua}}{{Mezcua}}{2019}]{Mezcua2019}
{Mezcua} M.,  2019, \mn@doi [Nature Astronomy] {10.1038/s41550-018-0662-2},
  \href {https://ui.adsabs.harvard.edu/abs/2019NatAs...3....6M} {3, 6}

\bibitem[\protect\citeauthoryear{{Noll}, {Burgarella}, {Giovannoli}, {Buat},
  {Marcillac}  \& {Mu{\~n}oz-Mateos}}{{Noll} et~al.}{2009}]{Noll2009}
{Noll} S.,  {Burgarella} D.,  {Giovannoli} E.,  {Buat} V.,  {Marcillac} D.,
  {Mu{\~n}oz-Mateos} J.~C.,  2009, \mn@doi [\aap]
  {10.1051/0004-6361/200912497}, \href
  {https://ui.adsabs.harvard.edu/abs/2009A&A...507.1793N} {507, 1793}

\bibitem[\protect\citeauthoryear{{Ochsenbein}, {Bauer}  \&
  {Marcout}}{{Ochsenbein} et~al.}{2000}]{Ochsenbein2000}
{Ochsenbein} F.,  {Bauer} P.,   {Marcout} J.,  2000, \mn@doi [\aaps]
  {10.1051/aas:2000169}, \href
  {https://ui.adsabs.harvard.edu/abs/2000A&AS..143...23O} {143, 23}

\bibitem[\protect\citeauthoryear{{Osterbrock} \& {Ferland}}{{Osterbrock} \&
  {Ferland}}{2006}]{Osterbrock2006}
{Osterbrock} D.~E.,  {Ferland} G.~J.,  2006, {Astrophysics of gaseous nebulae
  and active galactic nuclei}.
University Science Books

\bibitem[\protect\citeauthoryear{{Panessa}, {Bassani}, {Cappi}, {Dadina},
  {Barcons}, {Carrera}, {Ho}  \& {Iwasawa}}{{Panessa}
  et~al.}{2006}]{Panessa2006}
{Panessa} F.,  {Bassani} L.,  {Cappi} M.,  {Dadina} M.,  {Barcons} X.,
  {Carrera} F.~J.,  {Ho} L.~C.,   {Iwasawa} K.,  2006, \mn@doi [\aap]
  {10.1051/0004-6361:20064894}, \href
  {https://ui.adsabs.harvard.edu/abs/2006A&A...455..173P} {455, 173}

\bibitem[\protect\citeauthoryear{{Raichoor} et~al.,}{{Raichoor}
  et~al.}{2020}]{Raichoor2020}
{Raichoor} A.,  et~al., 2020, \mn@doi [Research Notes of the American
  Astronomical Society] {10.3847/2515-5172/abc078}, \href
  {https://ui.adsabs.harvard.edu/abs/2020RNAAS...4..180R} {4, 180}

\bibitem[\protect\citeauthoryear{{Reines} \& {Volonteri}}{{Reines} \&
  {Volonteri}}{2015}]{Reines2015}
{Reines} A.~E.,  {Volonteri} M.,  2015, \mn@doi [\apj]
  {10.1088/0004-637X/813/2/82}, \href
  {https://ui.adsabs.harvard.edu/abs/2015ApJ...813...82R} {813, 82}

\bibitem[\protect\citeauthoryear{{Reines}, {Greene}  \& {Geha}}{{Reines}
  et~al.}{2013}]{Reines2013}
{Reines} A.~E.,  {Greene} J.~E.,   {Geha} M.,  2013, \mn@doi [\apj]
  {10.1088/0004-637X/775/2/116}, \href
  {https://ui.adsabs.harvard.edu/abs/2013ApJ...775..116R} {775, 116}

\bibitem[\protect\citeauthoryear{{Reines}, {Plotkin}, {Russell}, {Mezcua},
  {Condon}, {Sivakoff}  \& {Johnson}}{{Reines} et~al.}{2014}]{Reines2014}
{Reines} A.~E.,  {Plotkin} R.~M.,  {Russell} T.~D.,  {Mezcua} M.,  {Condon}
  J.~J.,  {Sivakoff} G.~R.,   {Johnson} K.~E.,  2014, \mn@doi [\apjl]
  {10.1088/2041-8205/787/2/L30}, \href
  {https://ui.adsabs.harvard.edu/abs/2014ApJ...787L..30R} {787, L30}

\bibitem[\protect\citeauthoryear{{Shen}}{{Shen}}{2013}]{Shen2013}
{Shen} Y.,  2013, Bulletin of the Astronomical Society of India, \href
  {https://ui.adsabs.harvard.edu/abs/2013BASI...41...61S} {41, 61}

\bibitem[\protect\citeauthoryear{{Shen} et~al.,}{{Shen}
  et~al.}{2019}]{Shen2019}
{Shen} Y.,  et~al., 2019, \mn@doi [\apjs] {10.3847/1538-4365/ab074f}, \href
  {https://ui.adsabs.harvard.edu/abs/2019ApJS..241...34S} {241, 34}

\bibitem[\protect\citeauthoryear{{Simmonds}, {Bauer}, {Thuan}, {Izotov},
  {Stern}  \& {Harrison}}{{Simmonds} et~al.}{2016}]{Simmonds2016}
{Simmonds} C.,  {Bauer} F.~E.,  {Thuan} T.~X.,  {Izotov} Y.~I.,  {Stern} D.,
  {Harrison} F.~A.,  2016, \mn@doi [\aap] {10.1051/0004-6361/201629310}, \href
  {https://ui.adsabs.harvard.edu/abs/2016A&A...596A..64S} {596, A64}

\bibitem[\protect\citeauthoryear{{Skrutskie} et~al.,}{{Skrutskie}
  et~al.}{2006}]{Skrutskie2006}
{Skrutskie} M.~F.,  et~al., 2006, \mn@doi [\aj] {10.1086/498708}, \href
  {https://ui.adsabs.harvard.edu/abs/2006AJ....131.1163S} {131, 1163}

\bibitem[\protect\citeauthoryear{{Smith} et~al.,}{{Smith}
  et~al.}{2009}]{smith2009}
{Smith} N.,  et~al., 2009, \mn@doi [\apj] {10.1088/0004-637X/695/2/1334}, \href
  {https://ui.adsabs.harvard.edu/abs/2009ApJ...695.1334S} {695, 1334}

\bibitem[\protect\citeauthoryear{{Smith} et~al.,}{{Smith}
  et~al.}{2017}]{Smith2017}
{Smith} N.,  et~al., 2017, \mn@doi [\mnras] {10.1093/mnras/stw3204}, \href
  {https://ui.adsabs.harvard.edu/abs/2017MNRAS.466.3021S} {466, 3021}

\bibitem[\protect\citeauthoryear{{Stalevski}, {Fritz}, {Baes}, {Nakos}  \&
  {Popovi{\'c}}}{{Stalevski} et~al.}{2012}]{Stalevski2012}
{Stalevski} M.,  {Fritz} J.,  {Baes} M.,  {Nakos} T.,   {Popovi{\'c}}
  L.~{\v{C}}.,  2012, \mn@doi [\mnras] {10.1111/j.1365-2966.2011.19775.x},
  \href {https://ui.adsabs.harvard.edu/abs/2012MNRAS.420.2756S} {420, 2756}

\bibitem[\protect\citeauthoryear{{Stalevski}, {Ricci}, {Ueda}, {Lira}, {Fritz}
  \& {Baes}}{{Stalevski} et~al.}{2016}]{Stalevski2016}
{Stalevski} M.,  {Ricci} C.,  {Ueda} Y.,  {Lira} P.,  {Fritz} J.,   {Baes} M.,
  2016, \mn@doi [\mnras] {10.1093/mnras/stw444}, \href
  {https://ui.adsabs.harvard.edu/abs/2016MNRAS.458.2288S} {458, 2288}

\bibitem[\protect\citeauthoryear{{Takada} et~al.,}{{Takada}
  et~al.}{2014}]{Takada2014}
{Takada} M.,  et~al., 2014, \mn@doi [\pasj] {10.1093/pasj/pst019}, \href
  {https://ui.adsabs.harvard.edu/abs/2014PASJ...66R...1T} {66, R1}

\bibitem[\protect\citeauthoryear{{Volonteri}, {Lodato}  \&
  {Natarajan}}{{Volonteri} et~al.}{2008}]{Volonteri2008}
{Volonteri} M.,  {Lodato} G.,   {Natarajan} P.,  2008, \mn@doi [\mnras]
  {10.1111/j.1365-2966.2007.12589.x}, \href
  {https://ui.adsabs.harvard.edu/abs/2008MNRAS.383.1079V} {383, 1079}

\bibitem[\protect\citeauthoryear{{Wang} \& {Xu}}{{Wang} \&
  {Xu}}{2015}]{Wang2015}
{Wang} J.,  {Xu} D.~W.,  2015, \mn@doi [\aap] {10.1051/0004-6361/201424848},
  \href {https://ui.adsabs.harvard.edu/abs/2015A&A...573A..15W} {573, A15}

\bibitem[\protect\citeauthoryear{{Wright} et~al.,}{{Wright}
  et~al.}{2010}]{Wright2010}
{Wright} E.~L.,  et~al., 2010, \mn@doi [\aj] {10.1088/0004-6256/140/6/1868},
  \href {https://ui.adsabs.harvard.edu/abs/2010AJ....140.1868W} {140, 1868}

\bibitem[\protect\citeauthoryear{{Yang} et~al.,}{{Yang}
  et~al.}{2020}]{Yang2020}
{Yang} G.,  et~al., 2020, \mn@doi [\mnras] {10.1093/mnras/stz3001}, \href
  {https://ui.adsabs.harvard.edu/abs/2020MNRAS.491..740Y} {491, 740}

\bibitem[\protect\citeauthoryear{{York} et~al.,}{{York}
  et~al.}{2000}]{York2000}
{York} D.~G.,  et~al., 2000, \mn@doi [\aj] {10.1086/301513}, \href
  {https://ui.adsabs.harvard.edu/abs/2000AJ....120.1579Y} {120, 1579}

\makeatother
\end{thebibliography}




\appendix

\section{Spectral Energy Distribution Fitting}
\label{sec:sed}

\begin{figure*}
	\includegraphics[width=0.49\textwidth]{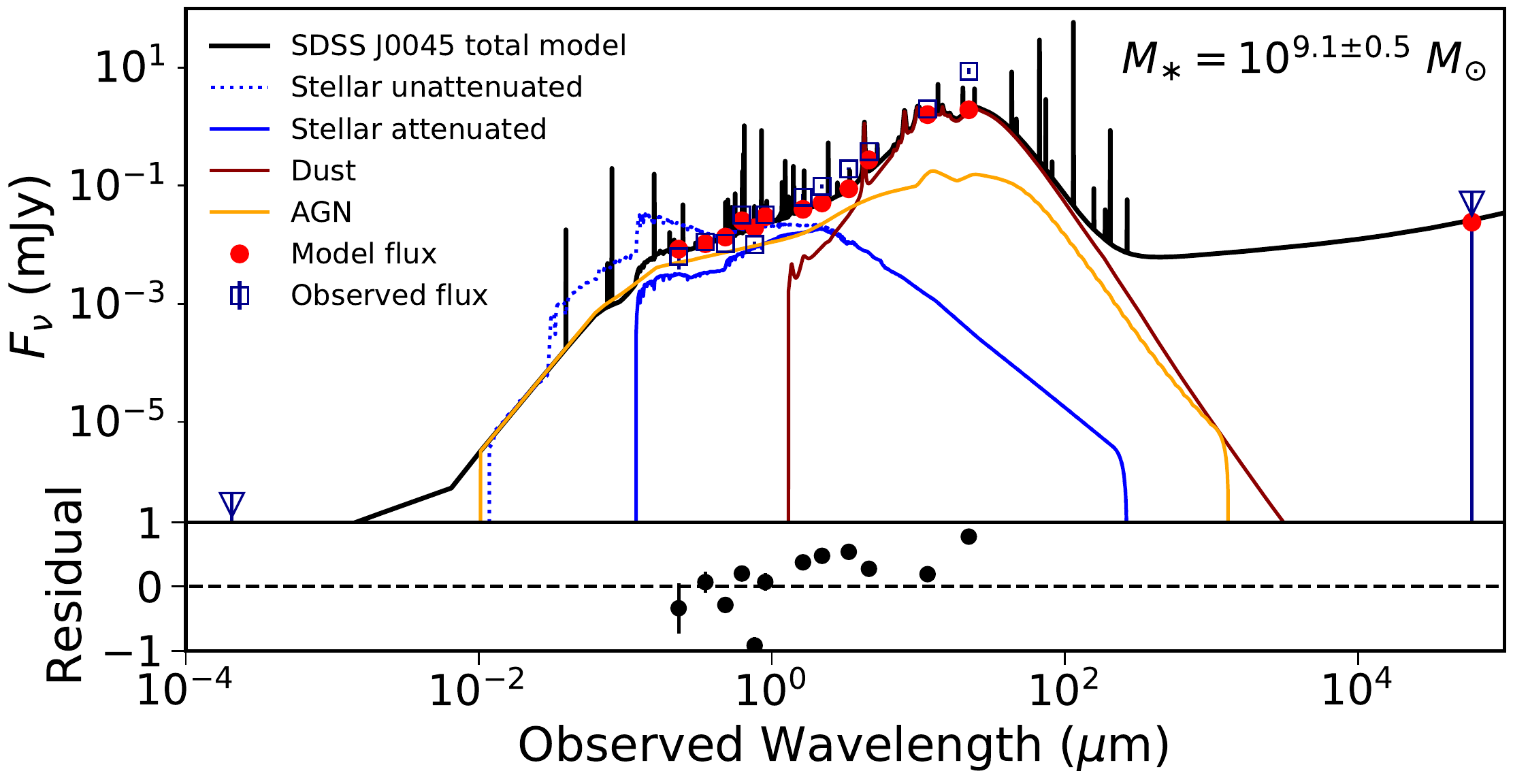}
	\includegraphics[width=0.49\textwidth]{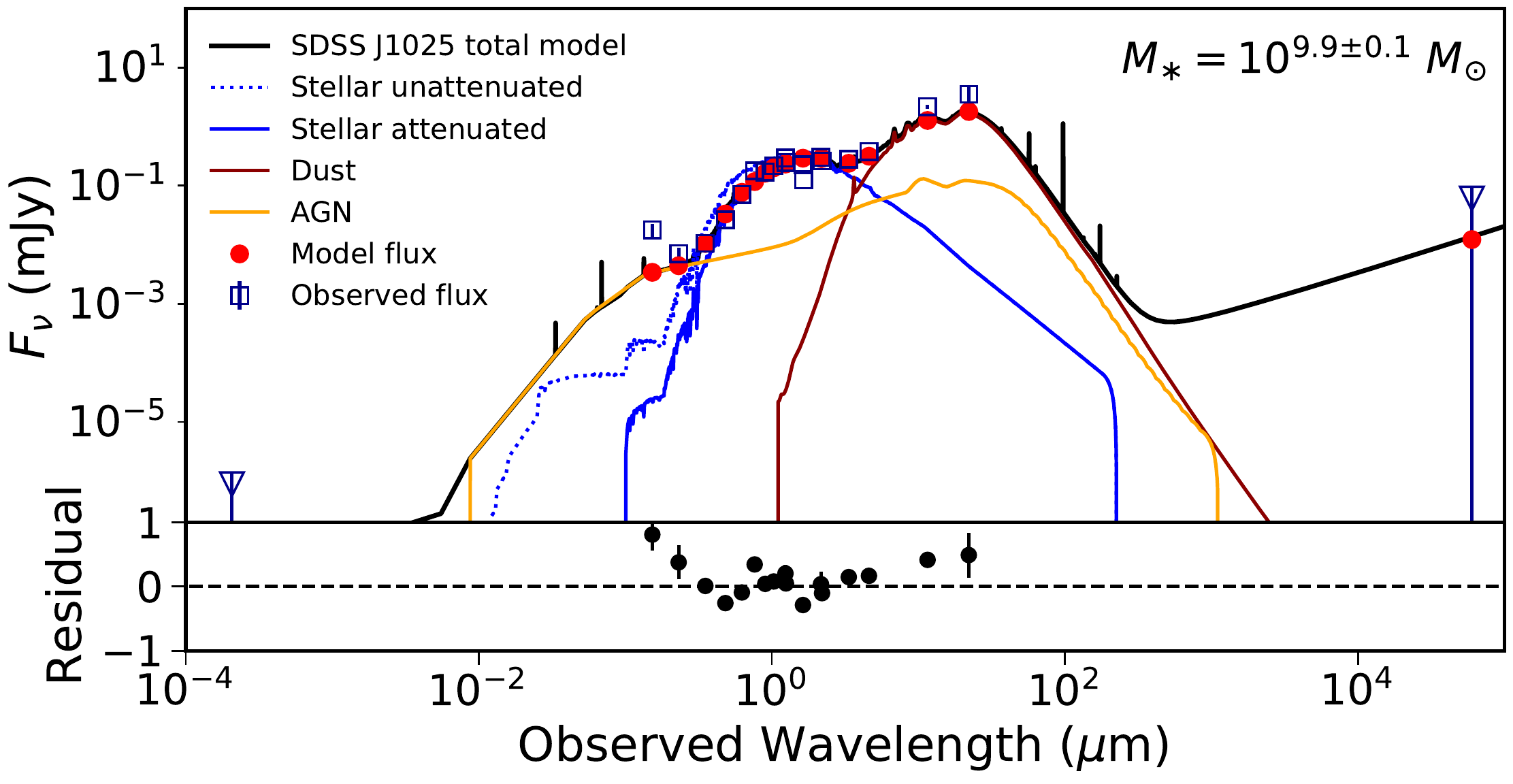}
	\includegraphics[width=0.49\textwidth]{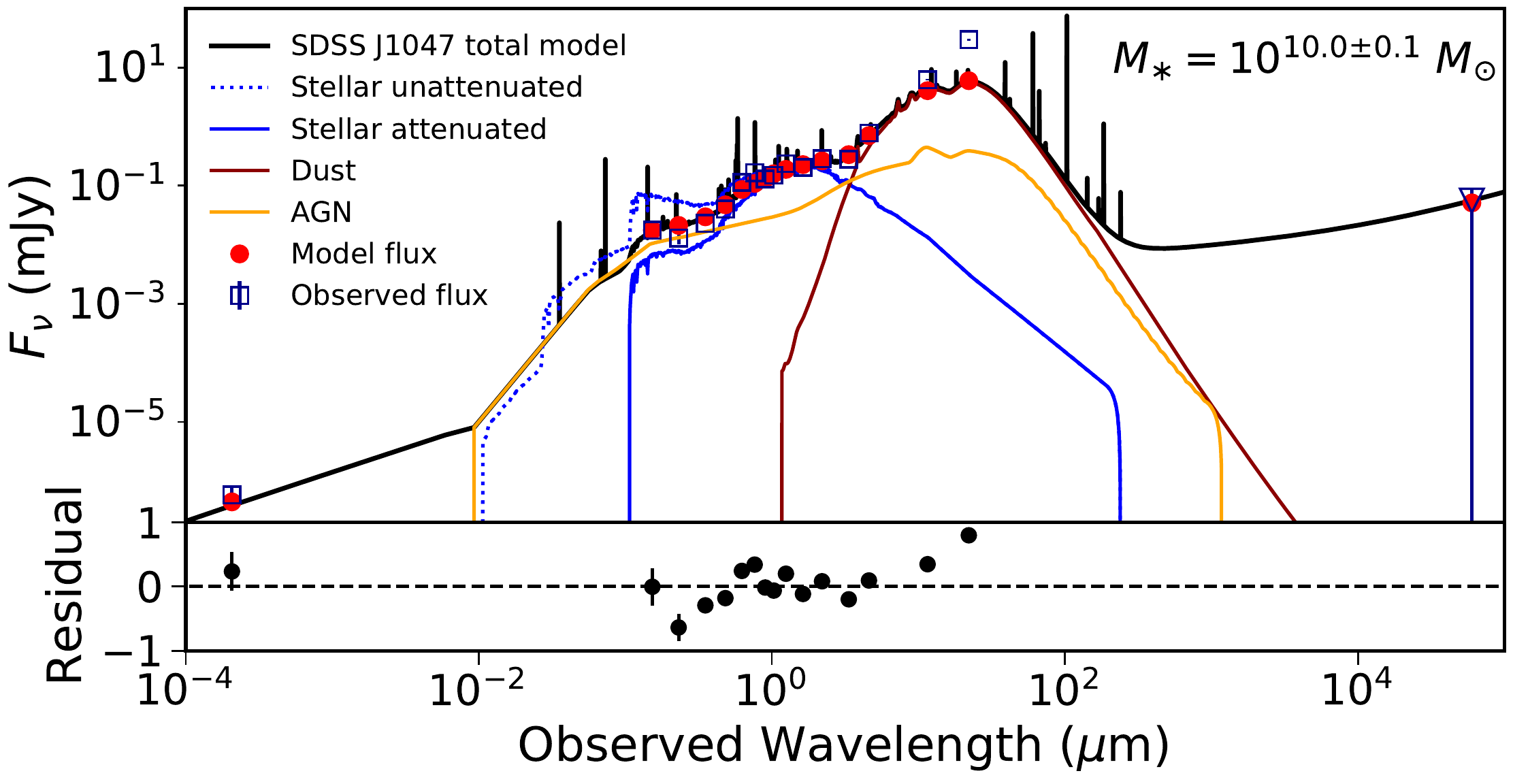}
	\includegraphics[width=0.49\textwidth]{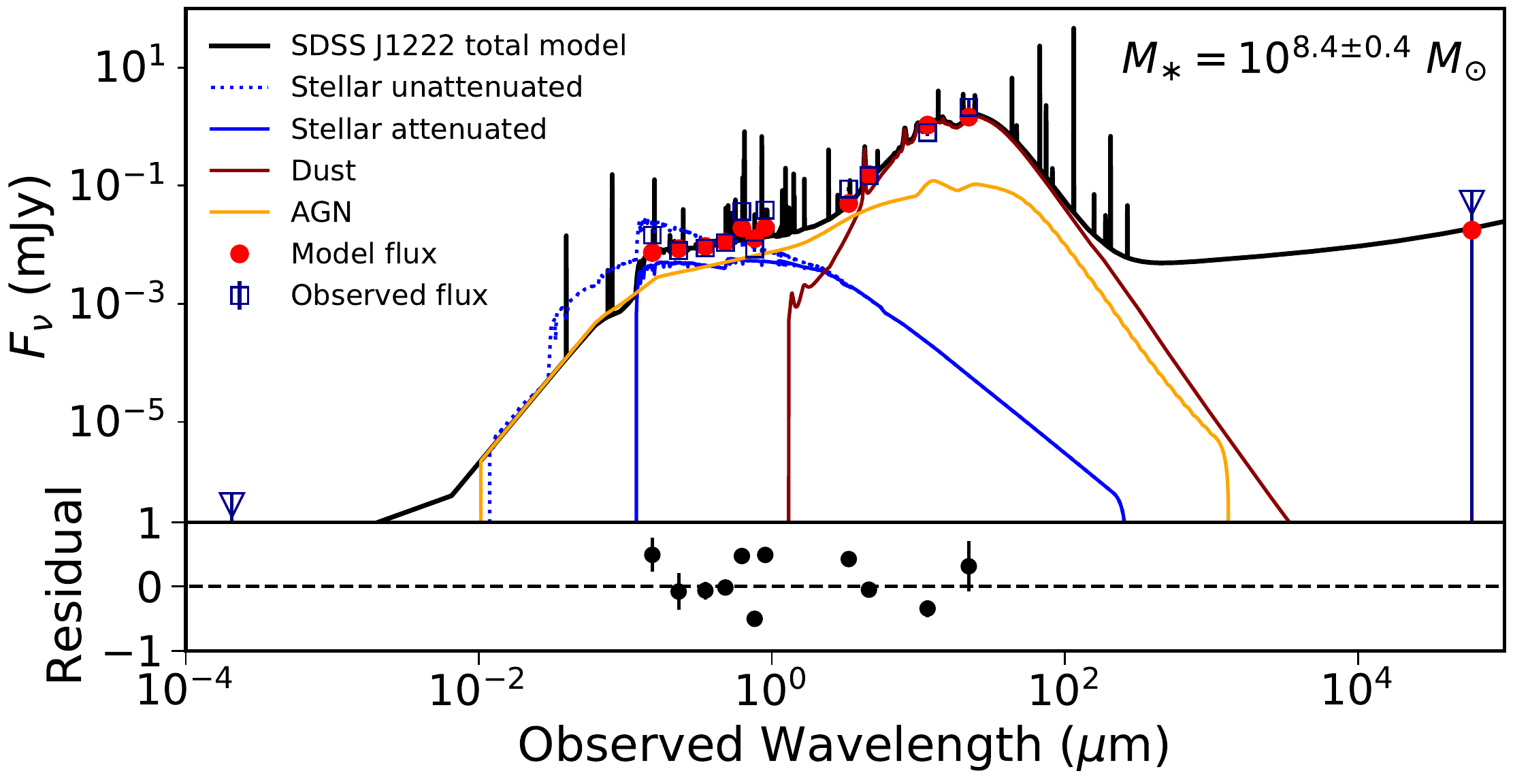}
    \caption{Fits to the broadband SEDs using \textsc{cigale}. The observed photometry from VizieR are shown as blue squares with upper-limits shown as inverted triangles. The model photometry are shown as red points. The best-fit model is shown in black. The components from attenuated stellar emission (blue dotted lines), attenuated stellar emission (blue solid lines), dust emission (dark red lines), and the AGN emission (orange) are also shown. Nebular emission is also fit, but not shown in the figure panels for clarity. All sources show a significant contribution from AGN emission. The estimated stellar mass and uncertainties from \textsc{cigale} are shown in the upper-right hand corner of each panel. Additional systematic uncertainties on the stellar mass may be up to 20 per cent \citep{Ciesla2015,Boquien2019}. The relative residual flux (observed - model) / observed are shown in at the bottom of each panel.}
    \label{fig:sed}
\end{figure*}

To estimate the host galaxy stellar masses, we model the multi-wavelength broadband spectral energy distribution (SED) for the four IT08 galaxies. For each source, we queried the VizieR \citep{Ochsenbein2000} SED database for available broadband photometry. We use available photometry from GALEX \citep{Martin2005}, SDSS \citep{York2000}, 2MASS \citep{Skrutskie2006}, UKIDDS \citep{Lawrence2007}, and WISE \citep{Wright2010}, as well as upper-limits (converted to 1$\sigma$) or detections from \emph{Chandra} and the VLA in Table~\ref{tab:prop}. When multi-epoch photometry are available, we take the mean value to quantify the average SED. We caution that SED variability driven by the AGN will introduce additional uncertainties in the SED model, particularly in the UV. However, given the low-levels of variability in Fig.~\ref{fig:lc}, this additional source of error is likely to be smaller than the final systematic uncertainty on the stellar mass estimate of up to 20 per cent \citep{Ciesla2015,Boquien2019} due to model choices and degeneracies \citep{Guo2020}. Finally, we caution that the error on the photometry may be under-estimated, particularly for the fainter sources, due to source confusion or sky background effects.

We use the \textsc{x-cigale} code \citep{Yang2020} to model the emission mechanisms and estimate the host galaxy stellar masses. \textsc{x-cigale} is an extension of \textsc{cigale} \citep{Boquien2019,Noll2009,Burgarella2005}, which works by imposing a self-consistent energy balance constraint between different emission and absorption mechanisms from the X-ray to radio. A large grid of models is computed and fitted to the data, allowing for an estimation of the star formation rate, stellar mass, and AGN contribution via a Bayesian-like analysis of the likelihood distribution.

We use a delayed exponential star formation history and vary the $e$-folding time and age of the stellar population assuming a sub-solar metallicity and the \citet{Chabrier2003} initial stellar mass function with the stellar population models of \citet{Bruzual2003}. We adopt the nebular emission template of \citet{Inoue2011}. We use the \citet{Leitherer2002} extension of the \citet{Calzetti2000} model for reddening due to dust extinction, and the \citet{Draine2014} updates to the \citet{Draine2007} model for dust emission. Finally, we adopt the SKIRTOR clumpy two-phase torus AGN emission model \citep{Stalevski2012,Stalevski2016} allowing for additional polar extinction. We allow the fractional AGN contribution to vary between 0.1 and 0.9 and a Type-1-like inclination angle varying from $i=10$ to $60$ degrees. The X-ray emission includes contribution from X-ray binaries. We allow the UV/X-ray correlation $\alpha_{\rm{ox}}$ to deviate up to 1 dex to account for X-ray weak nature of the AGN emission (although in this case the X-ray emission is not strongly constraining). The inferred stellar masses of the IT08 AGNs range from $\sim 10^{8.4} - 10^{10.0} M_{\odot}$. The best-fit SED model and stellar masses with uncertainties derived from \textsc{cigale} are shown in Fig.~\ref{fig:sed}. The stellar masses are broadly consistent with the local $M_{\bullet}-M_{\ast}$ from \citet{Reines2015}, which has a large intrinsic scatter of $\sim$0.55 dex, and predicts at black hole mass of $10^{6.4}\ M_{\odot}$ in a galaxy with stellar mass of $10^{10}\ M_{\odot}$.



\bsp	
\label{lastpage}
\end{document}